\RequirePackage{fix-cm}
\documentclass[twocolumn,epjc3]{svjour3}  
\smartqed  
\RequirePackage{graphicx}
\usepackage{multicol}
\RequirePackage{float}
\RequirePackage{hyperref}
\RequirePackage{amsmath}
\RequirePackage{amssymb}
\usepackage{widetext}
\RequirePackage{mathtools}      
\journalname{Eur. Phys. J. C}
\begin{document}

\title{Observational constraint on interacting Tsallis holographic dark energy in logarithmic Brans-Dicke theory}


\author{Y. Aditya\thanksref{e1,addr1}
        \and
        Sanjay Mandal\thanksref{e2,addr2}
        \and
        P.K. Sahoo\thanksref{e3,addr2} 
\and
D.R.K. Reddy\thanksref{e4,addr3}
      }

\thankstext{e1}{e-mail: yaditya2@gmail.com}
\thankstext{e2}{e-mail: sanjaymandal960@gmail.com}
\thankstext{e3}{e-mail: pksahoo@hyderabad.bits-pilani.ac.in}
\thankstext{e4}{e-mail: reddy\_ einstein@yahoo.com}


\institute{Department of Mathematics, GMR Institute of Technology, Rajam-532127, India \label{addr1}
           \and
           Department of Mathematics, Birla Institute of
Technology and Science-Pilani, Hyderabad Campus, Hyderabad-500078,
India \label{addr2}
          \and 
Department of Applied Mathematics, Andhra University, Visakhapatnam - 530003, India \label{addr3}
}

\date{Received: date / Accepted: date}

\maketitle

\begin{abstract}
In this paper, we investigate the dark energy phenomenon by studying the Tsallis holographic dark energy within the framework of Brans-Dicke (BD) scalar-tensor theory of gravity [Phys. Rev. \textbf{124}, 925 (1961)]. In this context, we choose the BD scalar field $\phi$ as a logarithmic function of the average scale factor $a(t)$ and Hubble horizon as the IR cutoff ($L=H^{-1}$). We reconstruct two cases of non-interacting and interacting fluid (dark sectors of cosmos) scenario. The physical behavior of the models are discussed with the help of graphical representation to explore the accelerated expansion of the universe. Moreover, the stability of the models are checked through squared sound speed $v_s^2$. The well-known cosmological plane i.e., $\omega_{de}-\omega^{\prime}_{de}$ is constructed for our models. We also include comparison of our findings of these dynamical parameters with observational constraints. It is also quite interesting to mention here that the results of deceleration, equation of state parameters and  $\omega_{de}-\omega^{\prime}_{de}$ plane coincide with the modern observational data.
\end{abstract}

\section{Introduction}
\hskip 0.6 cm Over the last few years, different cosmological observations such as type Ia supernovae \cite{per99,rie98}, cosmic microwave background radiation (CMBR) \cite{cal04,hau06}, Baryon Acoustic Oscillations (BAO) \cite{eis05,per10}, galaxy redshift survey \cite{fed09} and large scale structure \cite{koi06,dan08} strongly suggest that our universe is currently undergoing a phase of accelerated expansion.  Also, it is believed that the mysterious force known as dark energy (DE) with huge negative pressure is responsible for the current expanding universe with an acceleration. The present Planck data tells that there is 68.3\% DE of the total energy contents of the universe. We know very well that the standard cosmology has been extraordinarily successful but it unable to solve some serious issues including the search for the best DE candidate. Still there is some uncertainty in the origin and composition of DE except some particular ranges of the equation of state (EoS) parameter $\omega_{de}$. The approaches to answer this DE problem fall into two representative categories: one is to introduce dynamical DE in the right-hand side of the Einstein field equations in the framework of general relativity and the second one is to modify the left hand side of the Einstein equations, leading to a modified theories of gravity. In the literature, there are very nice reviews on both dynamical DE models and modified theories \cite{cop06}-\cite{noj17}. In the absence of any strong argument in favor of DE candidate, a variety of DE models have been discussed. Cosmological constant is the primary DE candidate for describing DE phenomenon but it has some serious problems like fine tuning and cosmic coincidence. Due to this reason, several dynamical DE models include a family of { scalar fields such as quintessence \cite{rat88}-\cite{zla99}, phantom \cite{cal02}-\cite{set07}, quintom \cite{fen05,cai10}, tachyon \cite{sen02}-\cite{set09}, K-essence \cite{arm00}, various Chaplygin gas models like generalized Chaplygin gas, extended Chaplygin gas and modified Chaplygin gas \cite{kam01}-\cite{sad16} have been developed.} 

The holographic DE (HDE) model \cite{li04} has been suggested in the context of quantum gravity with the help of holographic principle \cite{sus95}. This holographic principle says that the bound on the vacuum energy $\Lambda$ of a system with size $L$ should not cross the limit of the black hole (BH) mass having the same size due to the formation of BH in quantum field theory \cite{coh99}. { The energy density of HDE is defined as
\begin{equation}
\label{e1}\rho_{de}=3d^2m_p^2L^{-2}.
\end{equation}
where $m_p$ is the reduced Planck mass. This HDE model gives the relationship between the quantum fields energy density in vacuum to the various cut-offs such as infrared and ultraviolet. Granda and Oliveros \cite{gra08} proposed a IR cut-off containing the local quantities of Hubble and time derivative Hubble scales. The advantages of HDE with Granda and Oliveros cutoff (new HDE model) is that it depends on local quantities and avoids the causality problem appearing with event horizon IR cutoff. The new HDE model can also obtain the accelerated expansion of the universe and also showed that the transition redshift from deceleration phase ($q>0$) to acceleration phase ($q<0$) is consistent with current observational data \cite{gra08,dal08}. Nowadays, HDE attracted attention as it can alleviate the issue of cosmic coincidence, i.e., why the energy densities due the dark matter and the DE should have a constant ratio for the present universe \cite{pav07}. Also, various works show that the HDE model is in fairly good agreement with the observational data \cite{xu10}-\cite{dur11}. Nojiri and Odintsov \cite{noj06} have proposed unifying approach to early-time and late-time universe based on generalized HDE and phantom cosmology, and recently generalized this idea as Hinflation \cite{noj19a}. In recent years, various entropy formalism have been used to investigate the gravitational and cosmological setups. Also, some new HDE models are constructed such as Tsallis HDE (THDE) \cite{tav18,tsa13}, Renyi HDE model (RHDE) \cite{mor18} and Sharma-Mittal HDE (SMHDE) \cite{say18}. Among these models, RHDE based on the absence of interactions between cosmos sectors, and this model shows more stability by itself \cite{mor18}. SMHDE is classically stable in the case of non-interacting cosmos. The THDE model based on the Tsallis generalized entropy, which is never stable at the classical level \cite{tav18,tsa13,say18}. Hence, with this motivation, in this work we consider the HDE with another entropy formalism i.e., Tsallis HDE.}

As mentioned above, another approach to explore the present accelerated expansion of the universe is the modified theories of gravity. Standard Einstein's theory of gravitation may not completely explain the gravity at very high energy. In this situation cosmic acceleration would arise not from DE as a substance but rather from the dynamics of modified gravity \cite{tsu10}. The simplest alternative to general relativity is the scalar-tensor theory obtained by Brans and Dicke \cite{bra} (BD). But, in case of $w$ parameter value, there is a major difference between the theoretical and observational data. It is observed that the theoretical values are much less than that of observational value which motivates many researchers to explore various aspects of universe in BD framework \cite{ban01}-\cite{nai18}. Nojiri et al. \cite{noj05} have studied the properties of singularities in the phantom DE universe. They have, also, mentioned that the phantom-like behavior of EoS parameter $\omega_{de}$ may appear from BD theory, either from the non-minimal coupling of a scalar Lagrangian with gravity, or from negative (non-standard) potentials, or even the usual matter may appear in phantom-like nature. Recently, Saridakis et al. \cite{sar18} have discussed HDE through Tsallis entropy and its cosmological evolution through observational constraints. Barboza et al. \cite{bar15} and Nunes et al. \cite{nun16} have studied DE models through non-extensive Tsallis entropy framework and cosmological viability of non-gaussian statistics. Agostino \cite{ago19} has investigated the holographic principle through the nonadditive Tsallis entropy, used to describe the thermodynamic properties of nonstandard statistical systems such as the gravitational ones. Zadeh et al. \cite{zad} have explored the effects of different infrared cutoffs, such as the particle, Ricci horizons and the Granda-Oliveros cutoffs, on the properties of THDE model. Sharma and Pradhan \cite{sha19} have investigated diagnosing THDE models with statefinder and $\omega_{de}-\omega_{de}^\prime$ plane analysis. Sadri \cite{sadr19} has studied observational constraints on interacting THDE model. Sharif and Saba \cite{sharif19} have reconstructed the THDE model with Hubble horizon within the framework of $f(G,T)$ gravity. Nojiri et al. \cite{noj19} have studied modified cosmology from extended entropy with varying exponent.  Ghaffari et al. \cite{gaf} have discussed interacting and non-interacting THDE models by considering the Hubble horizon as the IR cutoff within BD scalar theory framework, while Jawad et al. \cite{jaw} have studied cosmological implications of THDE in modified version of BD scalar theory. In both the models the authors have considered the BD scalar field $\phi$ as a power function of average scale factor $a(t)$. Here, we are interested to extend the study of THDE models in BD theory with scalar field $\phi$ as logarithmic function of average scale factor.   

In this work, we are interested in studying the both non-interacting and interacting Tsallis holographic dark energy in Brans-Dicke scalar-tensor theory by considering homogeneous, isotropic FRW flat universe. Here, we focus our attention on the THDE models in BD theory with logarithmic expansion of average scale factor $a(t)$ for BD scalar field $\phi$. The present work is organized as follows: in Sect. 2, we derive BD field equations in the presence of THDE and pressure less dark matter (DM). Also, we have constructed non-interacting and interacting THDE models along with their complete physical discussion. Finally, in Sect. 3, we present the conclusions of this paper and also made a comparative analysis. 

\section{Tsallis holographic dark energy in BD theory}
\hskip 0.6cm We consider the homogeneous, isotropic and flat FRW metric in the form 
\begin{eqnarray}
\label{m1}ds^2=dt^2-a^2\bigg\{{dr^2}+r^2(d\theta^2+sin^2~\theta d\psi^2)\bigg\},
\end{eqnarray}
where $a(t)$ is the scale factor of the model. The spatial volume ($V$), Hubble parameter ($H$) and deceleration parameter ($q$) of this model are given by
\begin{eqnarray}
\label{m2}V&=&a^3\\
\label{m3}H&=&\frac{\dot{a}}{a}\\
\label{m4}q&=&-\frac{\dot{H}}{H^2}-1.
\end{eqnarray}

Different theories of gravitation have been proposed as alternatives to Einstein's general theory of gravity. But, the scalar-tensor theory formulated by Brans and Dicke \cite{bra} is supposed to be the best alternative to Einstein's theory. We consider the universe filled with pressure less DM with energy density $\rho_m$ and DE with density $\rho_{de}$. Hence, in this case the BD field equations for the combined scalar and tenor fields are given by 

\begin{eqnarray}
\label{b1}R_{ij}-\frac{1}{2}Rg_{ij}&=&-\frac{8\pi}{\phi}(T_{ij}+\overline{T}_{ij})-\phi^{-1}\left(\phi_{i;j}-g_{ij}\phi^{,k}_{;k}\right)\nonumber\\&&-w\phi^{-2}\left(\phi_{,i}\phi_{,j}-\frac{1}{2}g_{ij}\phi_{,k}\phi^{,k}\right),
\end{eqnarray}

\begin{equation}
\label{b2}\phi^{,k}_{;k}=\frac{8\pi}{(3+2w)}(T+\overline{T})
\end{equation}
and the energy conservation equation is
\begin{equation}
\label{b3}(T_{ij}+\overline{T}_{ij})_{;j}=0,
\end{equation}
which is a consequence of field equations \eqref{b1} and \eqref{b2}. Here, $R$ and $R_{ij}$ are the Ricci scalar and Ricci tensor respectively, $w$ is a dimensionless coupling constant. $T_{ij}$ and $\overline{T}_{ij}$ are energy-momentum tensors for pressure less DM and THDE, respectively, which are defined as
\begin{eqnarray}
\label{m5}T_{ij}&=&\rho_{m} u_iu_j\\
\label{m6}\overline{T}_{ij}&=&(\rho_{de}+p_{de})u_iu_j-p_{de}g_{ij}
\end{eqnarray}
where $p_{de}$ and $\rho_{de}$ are the pressure and energy density of DE respectively and $\rho_m$ is energy density of DM.

The field equations \eqref{b1} and \eqref{b2} for the metric \eqref{m1} are obtained as
\begin{eqnarray}
\label{m8}2\frac{\ddot{a}}{a}+\frac{\dot{a}^{2}}{a^{2}}+\frac{w}{2}\frac{\dot{\phi}^{2}}{\phi^{2}}+2\frac{\dot{a}\dot{\phi}}{a\phi}+\frac{\ddot{\phi}}{\phi}=-\frac{\omega_{de}\rho_{de}}{\phi}\\ \nonumber\\
\label{m9}3\frac{\dot{a}^{2}}{a^{2}}-\frac{w}{2}\frac{\dot{\phi}^{2}}{\phi^{2}}+3\frac{\dot{a}\dot{\phi}}{a\phi}=\frac{\rho_{de}+\rho_{m}}{\phi}\\ \nonumber\\
\label{m10}{\ddot{\phi}}+3\dot{\phi}\frac{\dot{a}}{a}=\frac{\rho_{de}(1-3\omega_{de})+\rho_{m}}{3+2w}
\end{eqnarray}
and the energy conservation equation \eqref{b3}, leads to
\begin{equation}
\label{e11}\dot{\rho}_{de}+\dot{\rho}_{m}+3H(\rho_{de}(1+\omega_{de})+\rho_m)=0,
\end{equation}
where overhead dot denotes ordinary differentiation with respect to time $t$. Here, $\omega_{de}$ is the equation of state (EoS) parameter of dark energy and is given by
\begin{eqnarray}
\label{e12}\omega_{de}=\frac{p_{de}}{\rho_{de}}.
\end{eqnarray}

In literature it is also common to use a power-law relation between BD scalar field $\phi$ and average scale factor \textquoteleft $a$' of the form $\phi=\phi_0a^l$  \cite{joh1,joh2}, where $\phi_0$ is a constant and $l$ is a power index. Various authors have discussed different aspects of this form of scalar field $\phi$ and have shown that it leads to constant deceleration parameter \cite{san16,sin12} and also time varying deceleration parameter \cite{shey,cha}. Recently, Kumar and Singh \cite{sinl1} have introduced a BD scalar field evolves as a logarithmic function of the average scale factor to study the evolution of holographic and new agegraphic DE models. The relation is given by 
\begin{eqnarray}
\label{m22c}\phi=\phi_1~ln(\beta_1+\beta_2~a(t))
\end{eqnarray}
where $\phi_1$, $\beta_1>1$ and $\beta_2>0$ are constants. Recently, Singh and Kumar \cite{sinl2}, Sadri and Vakili \cite{sad}, and Aditya and Reddy \cite{adi18} have investigated holographic DE models in BD theory using this logarithmic law for scalar field.

The energy density of Tsallis holographic DE model is given by \cite{tsa13}
\begin{eqnarray}
\label{e14}\rho_{de}=\eta L^{2\delta-4}
\end{eqnarray}
where $\eta$ is a parameter with dimensions $[L]^{-2\delta}$. For $\delta=1$ the above equation gives the usual holographic DE $\rho_{de}=3d^2m_p^2$, with $\eta=3d^2m_p^2$ and $d^2$ the model parameter. Also, it is interesting to mentioning that in the special case $\delta=2$ the above equation gives the standard cosmological constant model $\rho_{de}=constant=\Lambda$ (Saridakis et al. \cite{sar18}). 

By considering the Hubble horizon as the IR cutoff, $L =H^{-1}$, in BD theory the energy density \eqref{e14} takes the form 
\begin{eqnarray}
\label{e14a}\rho_{de}=3d^2\phi H^{-2\delta+4}.
\end{eqnarray}
The dimensionless density parameters are defined as
\begin{eqnarray}
\label{e12a}\Omega_{m}=\frac{\rho_{m}}{\rho_{cr}};~~\Omega_{de}=\frac{\rho_{de}}{\rho_{cr}};~~\Omega_{\phi}=\frac{\rho_{\phi}}{\rho_{cr}}
\end{eqnarray} 
where $\rho_{cr}=3m_p^2H^2$ is called the critical energy density and in BD theory it can be written as $\rho_{cr}=3\phi H^2$.

In the following sections, we consider the two cases: non-interacting model and interacting model. We determine, in both the cases energy density of THDE $\rho_{de}$, EoS parameter $\omega_{de}$, deceleration parameter $q$, squared sound speed $v_s^2$ and $\omega_{de}-\omega_{de}^\prime$ plane by solving the BD field equations. We also study their physical behavior.
\subsection{Non-interacting model}
First we consider that there is no energy exchange between the two fluids (cosmic sectors), and hence, the energy conservation equation \eqref{b3} leads us to the following separate conservation equations:
\begin{eqnarray}
\label{e13}\dot{\rho}_{de}+3H\rho_{de}(1+\omega_{de})=0,\\
\label{e13a}\dot{\rho}_{m}+3H\rho_m=0.
\end{eqnarray}
Taking differentiation with respect to time $t$ for Eq. \eqref{e13}, and using BD scalar field Eq. \eqref{m22c}, we have 

\begin{widetext}
\begin{eqnarray}
\label{e13ab}\dot{\rho}_{de}=\rho_{de}H\left(\frac{\beta_2a}{(\beta_1+\beta_2 a)~ln(\beta_1+\beta_2 a)}+(4-2\delta)\frac{\dot{H}}{H^2}\right). 
\end{eqnarray}
By taking the time derivative of Eq. \eqref{m9}, using Eqs. \eqref{m22c} and \eqref{e14a}-\eqref{e13a}, we obtain 

\begin{eqnarray}
\label{e16}\frac{\dot{H}}{H^2}&=&-\bigg\{\frac{6\beta_2 a}{(\beta_1+\beta_2 a)~ln(\beta_1+\beta_2 a)}-\frac{w\beta_2^2a^2}{(\beta_1+\beta_2 a)~[ln(\beta_1+\beta_2 a)]^2}+9\Omega_{de}(1+\omega_{de}+u)\nonumber\\&&-\frac{3\beta_2^2 a^2}{(\beta_1+\beta_2 a)^2~ln(\beta_1+\beta_2 a)}+\frac{w\beta_2^3a^3}{(\beta_1+\beta_2 a)^2~[ln(\beta_1+\beta_2 a)]^2}+\frac{w\beta_2^3a^3}{(\beta_1+\beta_2 a)^3~[\ln(\beta_1+\beta_2 a)]^3}\bigg\}\nonumber\\&&\times\bigg\{6+\frac{6\beta_2 a}{(\beta_1+\beta_2 a)~[ln(\beta_1+\beta_2 a)]}-\frac{w\beta_2^2a^2}{(\beta_1+\beta_2 a)~[ln(\beta_1+\beta_2 a)]^2}\bigg\}^{-1}
\end{eqnarray}
where $u=\frac{\rho_m}{\rho_{de}}=\frac{\Omega_m}{\Omega_{de}}$. On the other side, from equations \eqref{e14a} and \eqref{e13}, we obtain 

\begin{eqnarray}
\label{e15}\frac{\dot{H}}{H^2}=\frac{3}{2(\delta-2)}\left[1+\omega_{de}+\frac{\phi_1\beta_2 a}{3(\beta_1+\beta_2 a)~\ln(\beta_1+\beta_2 a)}\right].
\end{eqnarray}

From the above two Eqs. \eqref{e16} and \eqref{e15}, we find that

\begin{eqnarray}
\label{e17}\frac{\dot{H}}{H^2}&=&-\bigg\{\frac{3\beta_2 a(2+\Omega_{de})}{(\beta_1+\beta_2 a)~\ln(\beta_1+\beta_2 a)}-\frac{w\beta_2^2a^2}{(\beta_1+\beta_2 a)~[\ln(\beta_1+\beta_2 a)]^2}-\frac{3\beta_2^2 a^2}{(\beta_1+\beta_2 a)^2~\ln(\beta_1+\beta_2 a)}\nonumber\\&&+\frac{w\beta_2^3a^3}{(\beta_1+\beta_2 a)^2~[\ln(\beta_1+\beta_2 a)]^2}+\frac{w\beta_2^3a^3}{(\beta_1+\beta_2 a)^3~[\ln(\beta_1+\beta_2 a)]^3}+9\Omega_{de}u\bigg\}\nonumber\\&&\times\bigg\{6+6(\delta-2)\Omega_{de}+\frac{6\beta_2 a}{(\beta_1+\beta_2 a)~[\ln(\beta_1+\beta_2 a)]}-\frac{w\beta_2^2a^2}{(\beta_1+\beta_2 a)~[\ln(\beta_1+\beta_2 a)]^2}\bigg\}^{-1}
\end{eqnarray}
\end{widetext}
Taking time derivative of Eq. \eqref{e12a}, we obtain
\begin{eqnarray}
\label{e12b}\dot{\Omega}_{de}=2\Omega_{de} ~(1-\delta)~\frac{\dot{H}}{H}.
\end{eqnarray}
In order to observe the behavior density parameter of THDE, we define  $\Omega_{de}^\prime=\frac{\dot{\Omega}_{de}}{H}$, where the prime denotes derivative with respect to \textquoteleft$\ln~a(t)$'. Then from Eqs. \eqref{e17} and \eqref{e12b} we have

\begin{widetext}
\begin{eqnarray}
\label{e12c}\Omega_{de}^{\prime}&=&2\bigg\{\frac{3\beta_2 a(2+\Omega_{de})}{(\beta_1+\beta_2 a)~\ln(\beta_1+\beta_2 a)}-\frac{w\beta_2^2a^2}{(\beta_1+\beta_2 a)~[\ln(\beta_1+\beta_2 a)]^2}-\frac{3\beta_2^2 a^2}{(\beta_1+\beta_2 a)^2~\ln(\beta_1+\beta_2 a)}\nonumber\\&&+\frac{w\beta_2^3a^3}{(\beta_1+\beta_2 a)^2~[\ln(\beta_1+\beta_2 a)]^2}+\frac{w\beta_2^3a^3}{(\beta_1+\beta_2 a)^3~[\ln(\beta_1+\beta_2 a)]^3}+9\Omega_{de}u\bigg\}\times \Omega_{de}(\delta-1)\nonumber\\&&\times\bigg\{6+6(\delta-2)\Omega_{de}+\frac{6\beta_2 a}{(\beta_1+\beta_2 a)~[\ln(\beta_1+\beta_2 a)]}-\frac{w\beta_2^2a^2}{(\beta_1+\beta_2 a)~[\ln(\beta_1+\beta_2 a)]^2}\bigg\}^{-1}.
\end{eqnarray}
Also, from Eqs. \eqref{e16} and \eqref{e15} we obtain EoS parameter of THDE as
\begin{eqnarray}
	\label{e17a}\omega_{de}&=&-1-\Bigg\{\frac{6\beta_2 a}{(\beta_1+\beta_2 a)~\ln(\beta_1+\beta_2 a)}-\frac{w\beta_2^2a^2}{(\beta_1+\beta_2 a)~[\ln(\beta_1+\beta_2 a)]^2}+9\Omega_{de}u\nonumber\\&&+\frac{w\beta_2^3a^3}{(\beta_1+\beta_2 a)^2~[\ln(\beta_1+\beta_2 a)]^2}+\frac{w\beta_2^3a^3}{(\beta_1+\beta_2 a)^3~[\ln(\beta_1+\beta_2 a)]^3}-\frac{3\beta_2^2 a^2}{(\beta_1+\beta_2 a)^2~\ln(\beta_1+\beta_2 a)}\nonumber\\&&+\frac{\beta_2 a}{2(\delta-2)(\beta_1+\beta_2 a)~[\ln(\beta_1+\beta_2 a)]}\bigg\{6+\frac{6\beta_2 a}{(\beta_1+\beta_2 a)~[\ln(\beta_1+\beta_2 a)]}\nonumber\\&&-\frac{w\beta_2^2a^2}{(\beta_1+\beta_2 a)~[\ln(\beta_1+\beta_2 a)]^2}\bigg\}\Bigg\}\times \Bigg\{\frac{3}{2(\delta-2)}\bigg\{6+\frac{6\beta_2 a}{(\beta_1+\beta_2 a)~[\ln(\beta_1+\beta_2 a)]}\nonumber\\&&-\frac{w\beta_2^2a^2}{(\beta_1+\beta_2 a)~[\ln(\beta_1+\beta_2 a)]^2}+9\Omega_{de}\bigg\}\Bigg\}^{-1}.
	\end{eqnarray}
\end{widetext}

	\begin{figure}[H]
	\centering
	\includegraphics[width=1.1\linewidth]{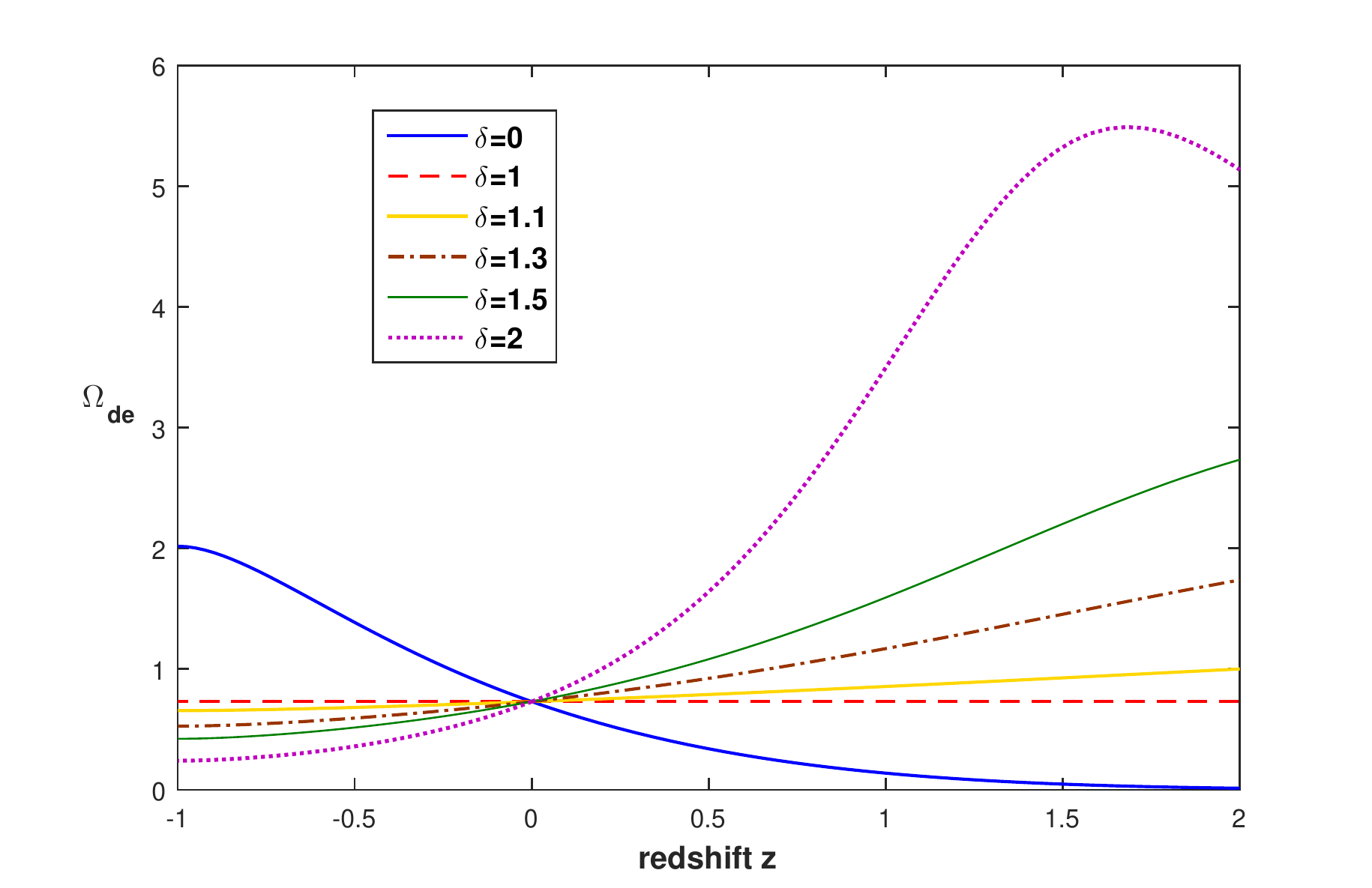}
	\caption{Plot of $\Omega_{de}$ of non-interacting THDE model versus redshift $z$ for $\beta_1=7.1$, $\beta_2=2.15$, $w=1000$, $\Omega^0_{de}=0.73$ and $u=0.3$.}
	\label{fig:om}
\end{figure}

\begin{figure}[H]
	\centering
	\includegraphics[width=1.1\linewidth]{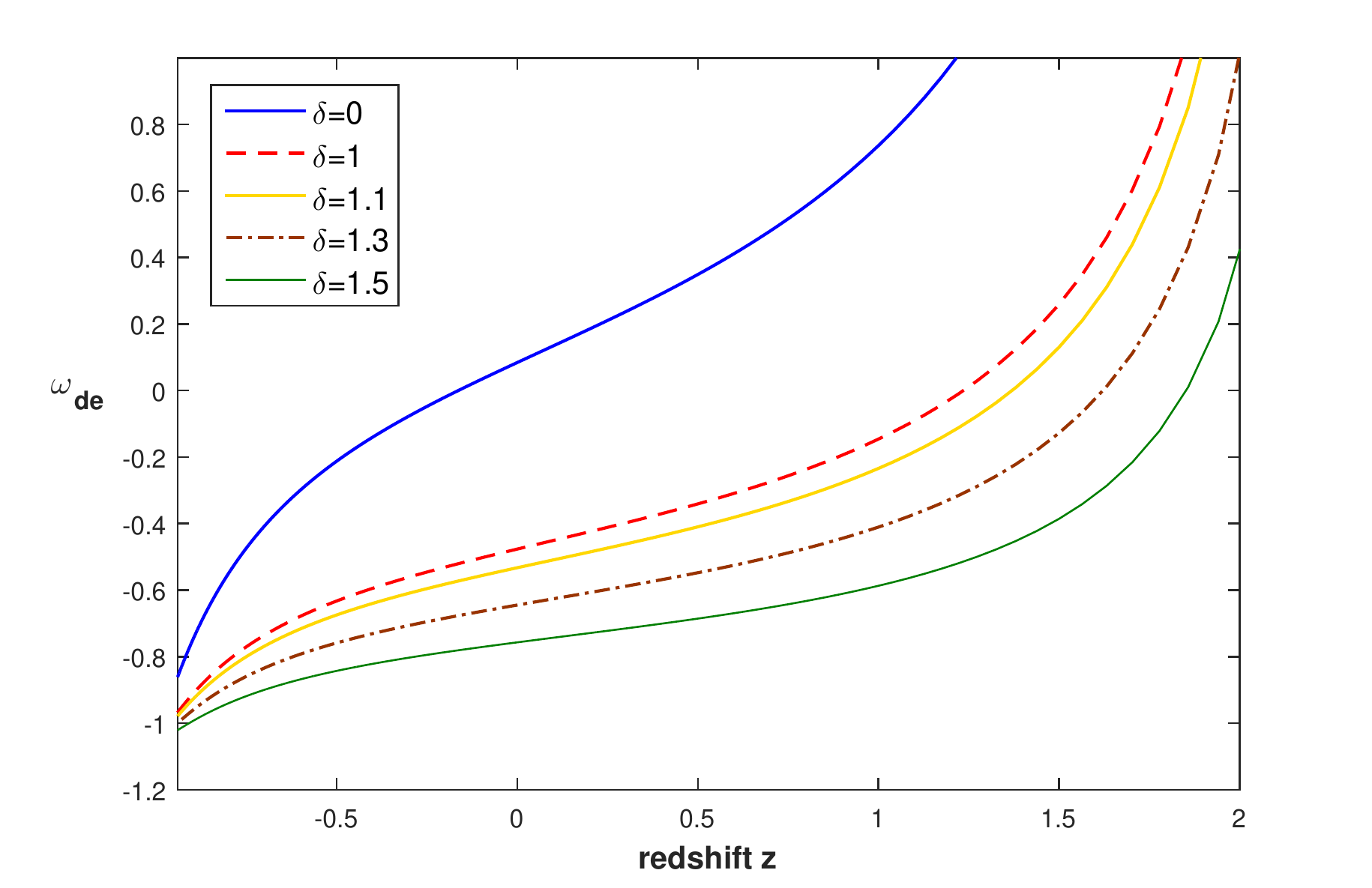}
	\caption{Plot of EoS parameter of non-interacting THDE model versus redshift $z$ for $\beta_1=7.1$, $\beta_2=2.15$, $w=1000$, $\Omega^0_{de}=0.73$ and $u=0.3$.}
	\label{fig:eos}
\end{figure}

To study the behavior of dimensionless density parameter of THDE $\Omega_{de}$, we solve the equation \eqref{e12c} numerically corresponding to redshift $z$ whose output is given in Fig. \ref{fig:om}. In the evolution of dynamical parameters the THDE parameter $\delta$ plays a crucial role and hence we have taken various values to $\delta=0,~1,~1.1,~1.3,$ $1.5,~2$. Here, we have used the initial value of $\Omega_{de}$ as $\Omega_{de}^0=0.73$ and additionally we have taken $\beta_1=7.1$, $\beta_2=2.15$, $w=1000$ \cite{gaf,acq} and $u=0.3$ \cite{gaf1}. It can be seen that the $\Omega_{de}$ is positive and decreasing function throughout the evolution, and finally it approaches to different positive constant values for $\delta=1,~1.1,~1.3,~1.5,~2$ whereas the density parameter is increasing function for $\delta=0$. Also, we observe that the steepness of density parameter $\Omega_{de}$ increases with increase in $\delta$. { Observations from Planck's data (2014) \cite{ade} have given the following constraints on the DE density parameter $\Omega_{de}=0.717^{+0.023}_{-0.020}$ (Planck +WP) and $\Omega_{de}=0.717^{+0.028}_{-0.024}$ (WMAP-9) and Planck's data (2018) \cite{agh} have given the constrains on DE density parameter as  $\Omega_{de}= 0.679 \pm 0.013$ (TT +lowE), $0.699 \pm 0.012$ (TE +lowE), $0.711^{+0.033}_{-0.026}$ (EE +lowE), $0.6834 \pm 0.0084$ (TT, TE, EE+lowE), $0.6847 \pm 0.0073$ (TT, TE, EE +lowE +lensing), $0.6889\pm 0.0056$ (TT, TE, EE +lowE +lensing +BAO) by implying various combination of observational schemes at 68\% confidence level. We observed that the THDE density parameter meets the above mentioned limits at present epoch, which shows that our results are consistent.} 

The EoS parameter is the relationship between pressure $p_{de}$ and energy density $\rho_{de}$ of DE whose expression is given by $\omega_{de}=\frac{p_{de}}{\rho_{de}}$. The EoS parameter is used to classify the decelerated and accelerated expansion of the universe and it categorizes various epochs as follows: when $\omega=1$, it represents stiff fluid, if $\omega=1/3$, the model shows the radiation dominated phase while $\omega=0$ represents matter dominated phase. In DE dominated accelerated phase, $-1<\omega<-1/3$ shows the quintessence phase and $\omega=-1$ shows the cosmological constant, i.e., $\Lambda$CDM model and $\omega<-1$ yields the phantom era. In Fig. \ref{fig:eos}, we have plotted EoS parameter versus redshift $z$ for different values of $\delta$. { We observe from Fig. \ref{fig:eos} that the EoS parameter starts from matter dominated era, then goes towards quintessence DE era and finally approaches to vacuum DE era for different values of $\delta=0,~1,~1.1,$ $1.3,~1.5$. For $\delta=2$, it is clear from the equation \eqref{e17a} that the EoS parameter becomes $-1$, i.e., cosmological constant. This is in agreement with the results obtained by Saridakis et al. \cite{sar18}. Also, it is interesting to mention here that as $\delta$ decreases the transition from matter dominated phase to dark energy phase is delayed considerably.} It may be noted that the EoS parameter, in this non-interacting case, never crosses the phantom divided line (PDL) $\omega_{de}=-1$.  

Deceleration parameter is another important cosmological parameter by means of which one can distinguish current accelerated or early decelerated expansions. It is defined as 
\begin{eqnarray}
\label{e18}q=-1-\frac{\dot{H}}{H^2}.
\end{eqnarray}
In this case, with the help of Eq. \eqref{e17} this deceleration parameter can be obtained as

\begin{widetext}
\begin{eqnarray}
\label{e19}q&=&-1+\bigg\{\frac{3\beta_2 a(2+\Omega_{de})}{(\beta_1+\beta_2 a)~\ln(\beta_1+\beta_2 a)}-\frac{3\beta_2^2 a^2}{(\beta_1+\beta_2 a)^2~\ln(\beta_1+\beta_2 a)}-\frac{w\beta_2^2a^2}{(\beta_1+\beta_2 a)~[\ln(\beta_1+\beta_2 a)]^2}\nonumber\\&&+\frac{w\beta_2^3a^3}{(\beta_1+\beta_2 a)^2~[\ln(\beta_1+\beta_2 a)]^2}+\frac{w\beta_2^3a^3}{(\beta_1+\beta_2 a)^3~[\ln(\beta_1+\beta_2 a)]^3}+9\Omega_{de}u\bigg\}\nonumber\\&&\times\bigg\{6+6(\delta-2)\Omega_{de}+\frac{6\beta_2 a}{(\beta_1+\beta_2 a)~[\ln(\beta_1+\beta_2 a)]}-\frac{w\beta_2^2a^2}{(\beta_1+\beta_2 a)~[\ln(\beta_1+\beta_2 a)]^2}\bigg\}^{-1}
\end{eqnarray}
\end{widetext}

\begin{figure}[H]
	\centering
	\includegraphics[width=1.1\linewidth]{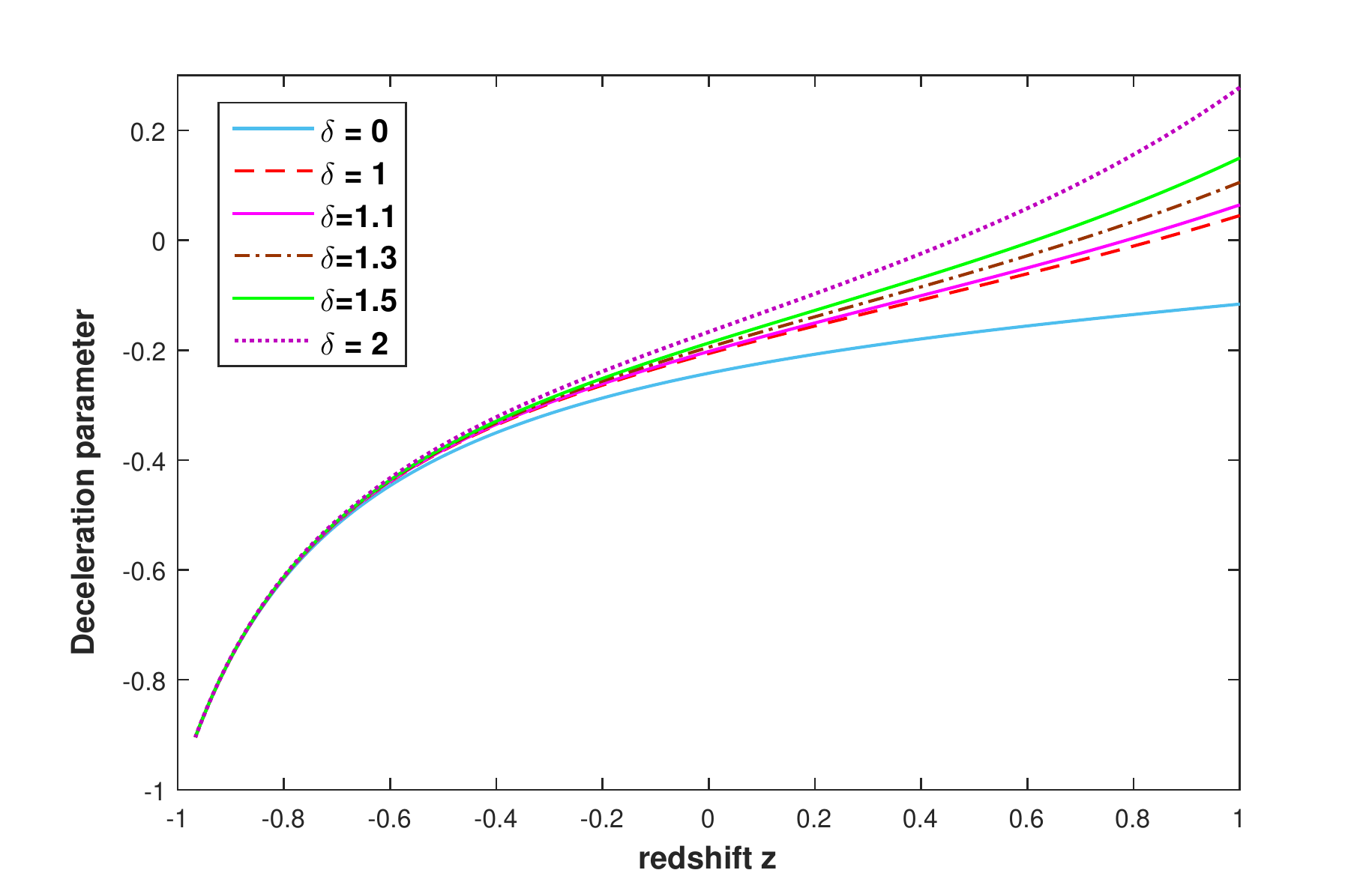}
	\caption{Plot of deceleration parameter of non-interacting THDE model versus redshift $z$ for $\beta_1=7.1$, $\beta_2=2.15$, $w=1000$, $\Omega^0_{de}=0.73$ and $u=0.3$.}
	\label{fig:nq}
\end{figure}

The plot of above constructed deceleration parameter versus redshift $z$ is shown in Fig. \ref{fig:nq} for different values of $\delta$ and assumed values of other parameters. { It may be noted that for $\delta=1,~1.1,~1.3,~1.5,~2$ the model exhibits a smooth transition from early deceleration era to the current acceleration era of the universe, whereas for $\delta=0$ the model remains in the accelerated phase.} Moreover, the transition redshift $z_t$ is decreasing as $\delta$ increases. We observed that the transition redshift $z_t$ from a deceleration phase to an accelerated universe lies within the interval $0.57<z_t<0.77$. This is in accordance with the recent cosmological observations \cite{gio,kom}. 

\begin{figure}[!ht]
	\centering
	\includegraphics[width=0.95\linewidth]{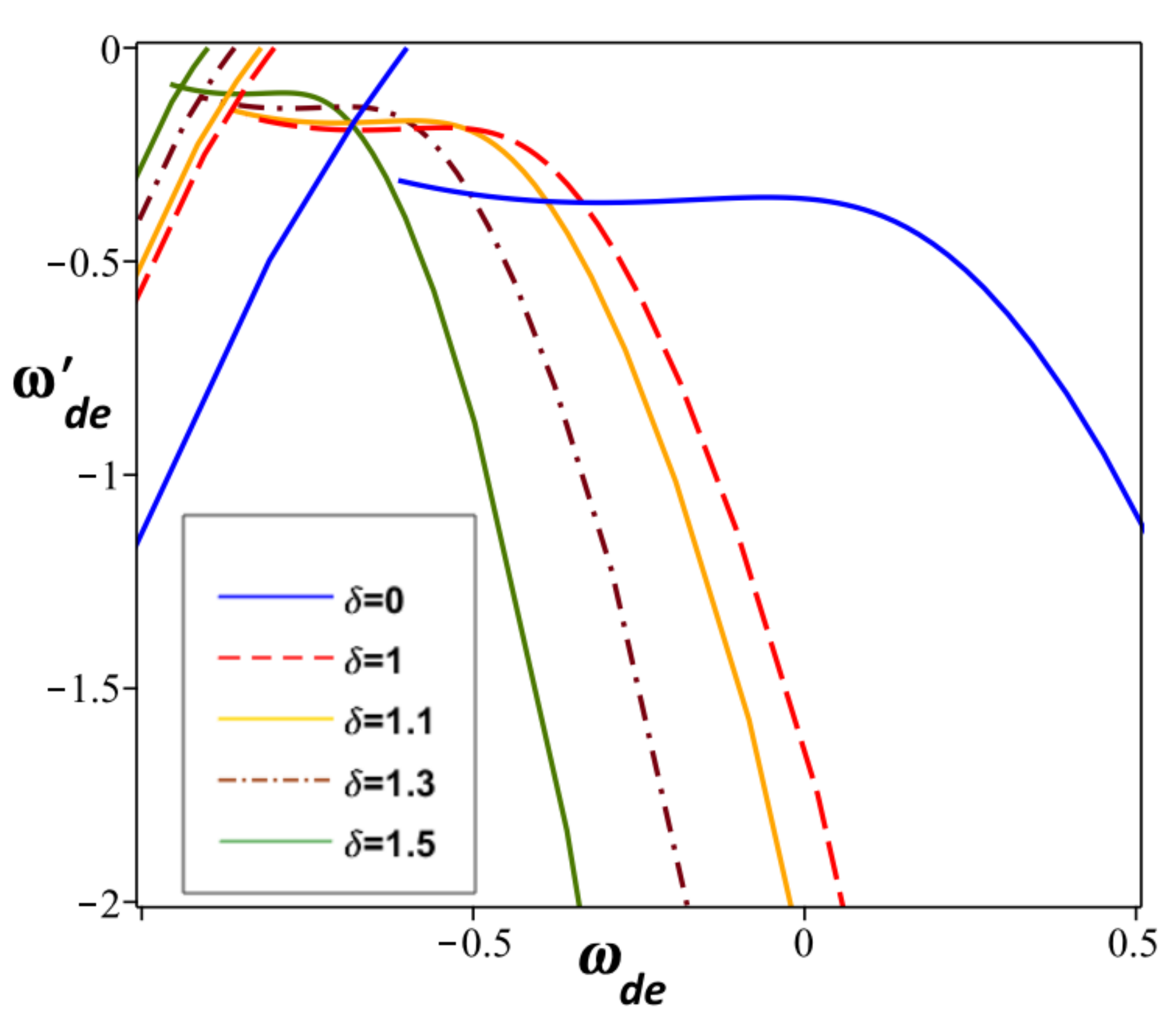}
	\caption{Plot of $\omega_{de}-\omega_{de}^\prime$ plane of non-interacting THDE model for $\beta_1=7.1$, $\beta_2=2.15$, $w=1000$, $\Omega^0_{de}=0.73$ and $u=0.3$.}
	\label{fig:eosp}
\end{figure}

The $\omega_{de}-\omega_{de}^\prime$ (where $\prime$ denotes differentiation with respect to $\ln~a$) plane explains the accelerated expansion regions of the universe. Caldwell and Linder \cite{cal} have firstly proposed this plane for analyzing the quintessence scalar field. Two different classes have been characterized from this plane known as thawing region ($\omega_{de}^\prime>0$ when $\omega_{de}<0$) and freezing region ($\omega_{de}^\prime<0$ when $\omega_{de}<0$) on the $\omega_{de}^\prime-\omega_{de}$ plane. Here, for non-interacting model, we have developed $\omega_{de}-\omega_{de}^\prime$ plane for different values of $\delta$ as given in Fig. \ref{fig:eosp}. From Fig. \ref{fig:eosp}, it can be observed that $\omega_{de}-\omega_{de}^\prime$ plane corresponds to freezing region for all the values of $\delta=0$,~$1$,~$1.1$,~$1.3$,~$1.5$. Observational data says that the expansion of the universe is comparatively more accelerating in freezing region. { Also, for $\delta=2$ the model coincides with the $\Lambda CDM$ limit $(\omega_{de},\omega_{de}^\prime)=(-1,0)$}. Hence, the behavior of $\omega_{de}-\omega_{de}^\prime$ plane is consistent with the present day observations. 

\begin{figure}[!ht]
	\centering
	\includegraphics[width=0.8\linewidth]{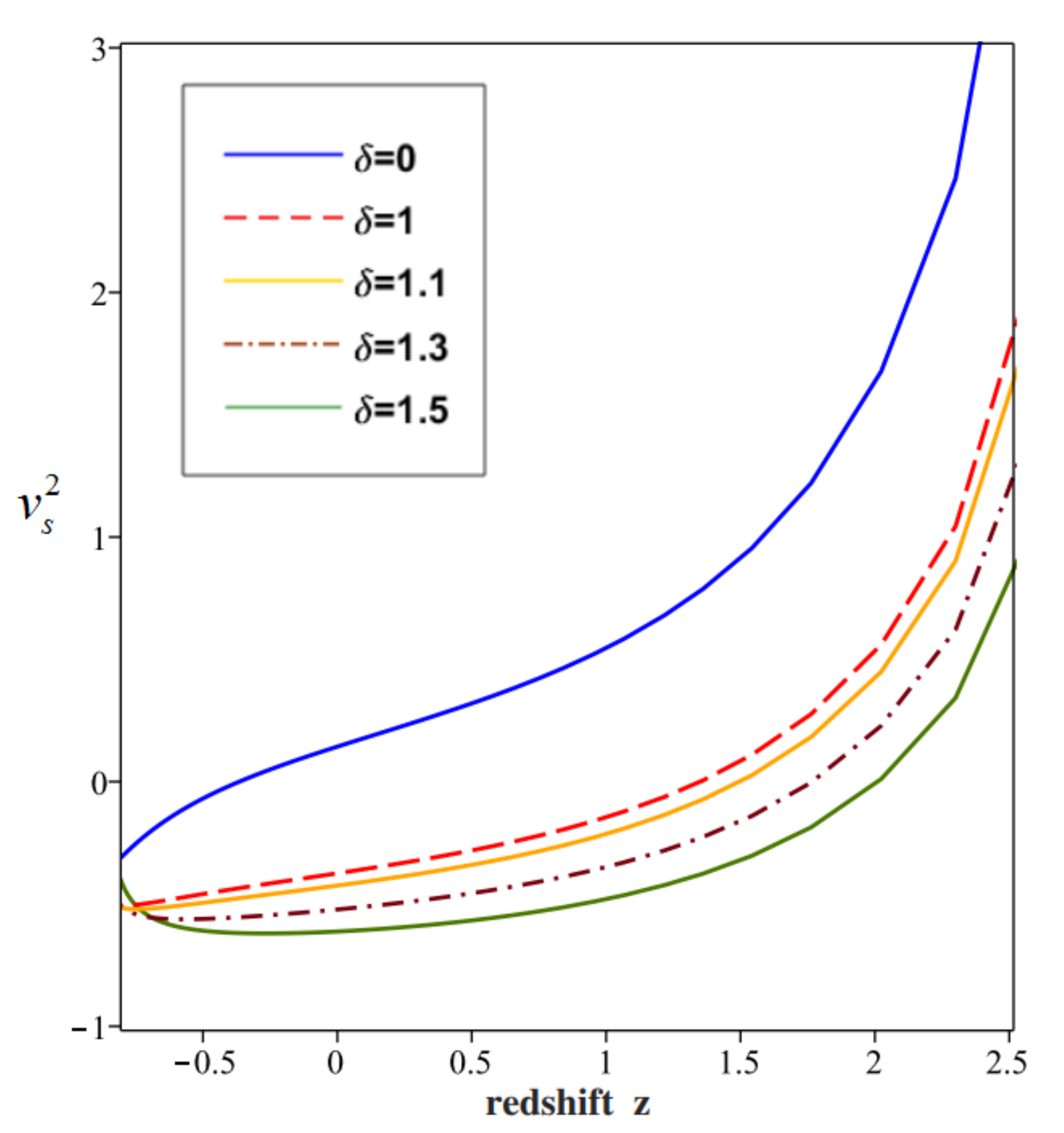}
	\caption{Plot of squared sound speed of non-interacting THDE model versus redshift $z$ for $\beta_1=7.1$, $\beta_2=2.15$, $w=1000$, $\Omega^0_{de}=0.73$ and $u=0.3$.}
	\label{fig:vs}
\end{figure}

In order to check the stability against small perturbations of our non interacting THDE model in this scenario, we obtain the squared speed of sound. Sign of square of speed of sound plays a crucial role, i.e., its negativity ($v_s^2<0$) represents instability and vice versa. It can be defined as follows
\begin{equation}
\label{vs}v_s^2=\frac{\dot{p}_{de}}{\dot{\rho}_{de}}.
\end{equation}
By differentiating the relation $\omega_{de}=\frac{p_{de}}{\rho_{de}}$ with respect to time $t$ and dividing by $\dot{\rho}_{de}$, we get
\begin{equation}
\label{vs1}v_s^2=\omega_{de}+\frac{\rho_{de}}{\dot{\rho}_{de}}~\dot{\omega}_{de}
\end{equation}

Now, using Eq. \eqref{e13ab} in above Eq. \eqref{vs1}, we get 
\begin{eqnarray}
\label{vs2}v_s^2=\omega_{de}+\frac{\dot{\omega}_{de}}{H\left(\frac{\beta_2a}{(\beta_1+\beta_2 a)~\ln(\beta_1+\beta_2 a)}+(4-2\delta)\frac{\dot{H}}{H^2}\right)}
\end{eqnarray}
where $\omega_{de}$ and $\frac{\dot{H}}{H^2}$ are, respectively, given in Eqs. \eqref{e17} and \eqref{e17a}. 
  
In the present scenario, we develop the squared speed of sound trajectories in terms redshift $z$ as shown in Fig. \ref{fig:vs} using same values of the constants in the model. For all the values of $\delta$, we can observe from Fig. \ref{fig:vs} that initially $v_s^2$ exhibits a decreasing behavior but with positive sign. This shows that the model stable at early epochs. Later, all the curves related to the parameter $\delta$ exhibit negative behavior. Thus for the non-interacting THDE model is found the stability at initial epoch and then becomes unstable in later epochs.

\subsection{Interacting Model}                  
Here, we consider the interaction between the two fluids. { Since the nature of both DE and dark matter is still unknown, there is no physical argument to exclude the possible interaction between them. Some observational evidences of the interaction in dark sectors have been found recently \cite{ber07,ber09}. Abdalla et al. \cite{abd09,abd10} have investigated the signature of interaction between DE and dark matter by using optical, X-ray and weak lensing data from the relaxed galaxy clusters. So, it is reasonable to assume the interaction between DE and dark matter in cosmology.}

 For this purpose we can write the energy conservation equations as
\begin{eqnarray}
\label{e13b}\dot{\rho}_{de}+3H\rho_{de}(1+\omega_{de})=-Q,\\
\label{e13ba}\dot{\rho}_{m}+3H\rho_m=Q.
\end{eqnarray}
where the quantity $Q$ represents interaction between DE components. { From the equations \eqref{e13b} and \eqref{e13ba} we can say that the total energy is conserved i.e., $\dot{\rho}_{tot}+3H(\omega_{tot}+1)\rho_{tot}=0$, where $\rho_{tot}=\rho_m + \rho_{de}$. Since there is no natural information from fundamental physics on the interaction term $Q$, one can only study it to a phenomenological level. Various forms of interaction term extensively considered in literature include $Q=3cH\rho_m$, $Q=3cH\rho_{de}$ and $Q=3cH(\rho_m+\rho_{de})$. Here, $c$ is a coupling constant and positive $c$ means that DE decays into dark matter, while negative $c$ means dark matter decays into DE. Also, there are many other forms for interaction term considered in literature which are defined as $Q=\gamma\dot{\rho}_i$ and $Q=3cH\gamma\rho_i+\gamma\dot{\rho}_i$ where $i={m, de, tot}$. A detailed analysis of linear interacting terms can be found in Izaquierdo and Pavon \cite{izq10}, Ferreira et al. \cite{fer13}, Sadeghi et al. \cite{sad15,sad16} and Wei \cite{wei}. Most of these interactions are either positive or negative and can not give the possibility to change their signs. Cai and Su \cite{cai10a} have proposed a new sign-changeable interaction by allowing $\dot{\rho}$ and deceleration parameter$q$ into interacting term as $Q=q(\gamma\dot{\rho}+3cH\rho_i)$ where $i={m, de, tot}$. Here $\gamma$ and $c$ dimensionless constants and deceleration parameter is defined in equation \eqref{m4}. Recently, Sadri et al. \cite{sadr19a} have compared different phenomenological linear as well as non-linear interaction cases in the framework of the holographic Ricci DE model and they found that the linear interaction are the best cases among the others. Inspired by the works of Pavon and Zimdahl \cite{pav05}, Sadeghi et al. \cite{sad13}, Honarvaryan et al. \cite{hon}, Zadeh et al. \cite{zad} and Sadri \cite{sadr19}, in this work, we consider the linear interacting term as $Q=3c^2H(\rho_m+\rho_{de})$. It is worthwhile to mention here that this work can be extended using sign-changeable interaction term, which will be done in other forthcoming articles.}

Now, combining the time derivative of Eq. \eqref{m9} with Eqs. \eqref{m22c}-\eqref{e12a}, \eqref{e13b} and \eqref{e13ba}, we easily get 

\begin{widetext}
\begin{eqnarray}
\label{e20}\frac{\dot{H}}{H^2}&=&-\bigg\{\frac{3\beta_2 a(2-\Omega_{de})}{(\beta_1+\beta_2 a)~\ln(\beta_1+\beta_2 a)}-\frac{w\beta_2^2a^2}{(\beta_1+\beta_2 a)~[\ln(\beta_1+\beta_2 a)]^2}-\frac{3\beta_2^2 a^2}{(\beta_1+\beta_2 a)^2~\ln(\beta_1+\beta_2 a)}\nonumber\\&&+\frac{w\beta_2^3a^3}{(\beta_1+\beta_2 a)^2~[\ln(\beta_1+\beta_2 a)]^2}+\frac{w\beta_2^3a^3}{(\beta_1+\beta_2 a)^3~[\ln(\beta_1+\beta_2 a)]^3}-9\Omega_{de}\left(u(c^2-1)+c^2\right)\bigg\}\nonumber\\&&\times\bigg\{6+6(\delta-2)\Omega_{de}+\frac{6\beta_2 a}{(\beta_1+\beta_2 a)~[\ln(\beta_1+\beta_2 a)]}-\frac{w\beta_2^2a^2}{(\beta_1+\beta_2 a)~[\ln(\beta_1+\beta_2 a)]^2}\bigg\}^{-1}
\end{eqnarray}

In view of Eq. \eqref{e12b}, from Eq. \eqref{e20} we obtain 

\begin{eqnarray}
\label{e20a}\Omega_{de}^\prime&=&2\Omega_{de}(\delta-1)\bigg\{\frac{3\beta_2 a(2-\Omega_{de})}{(\beta_1+\beta_2 a)~\ln(\beta_1+\beta_2 a)}-\frac{w\beta_2^2a^2}{(\beta_1+\beta_2 a)~[\ln(\beta_1+\beta_2 a)]^2}-\frac{3\beta_2^2 a^2}{(\beta_1+\beta_2 a)^2~\ln(\beta_1+\beta_2 a)}\nonumber\\&&+\frac{w\beta_2^3a^3}{(\beta_1+\beta_2 a)^2~[\ln(\beta_1+\beta_2 a)]^2}+\frac{w\beta_2^3a^3}{(\beta_1+\beta_2 a)^3~[\ln(\beta_1+\beta_2 a)]^3}-9\Omega_{de}\left(u(c^2-1)+c^2\right)\bigg\}\nonumber\\&&\times\bigg\{6+6(\delta-2)\Omega_{de}+\frac{6\beta_2 a}{(\beta_1+\beta_2 a)~[\ln(\beta_1+\beta_2 a)]}-\frac{w\beta_2^2a^2}{(\beta_1+\beta_2 a)~[\ln(\beta_1+\beta_2 a)]^2}\bigg\}^{-1}
\end{eqnarray}
\end{widetext}

\begin{figure*}
\begin{multicols}{2}
    \includegraphics[width=1.1\linewidth]{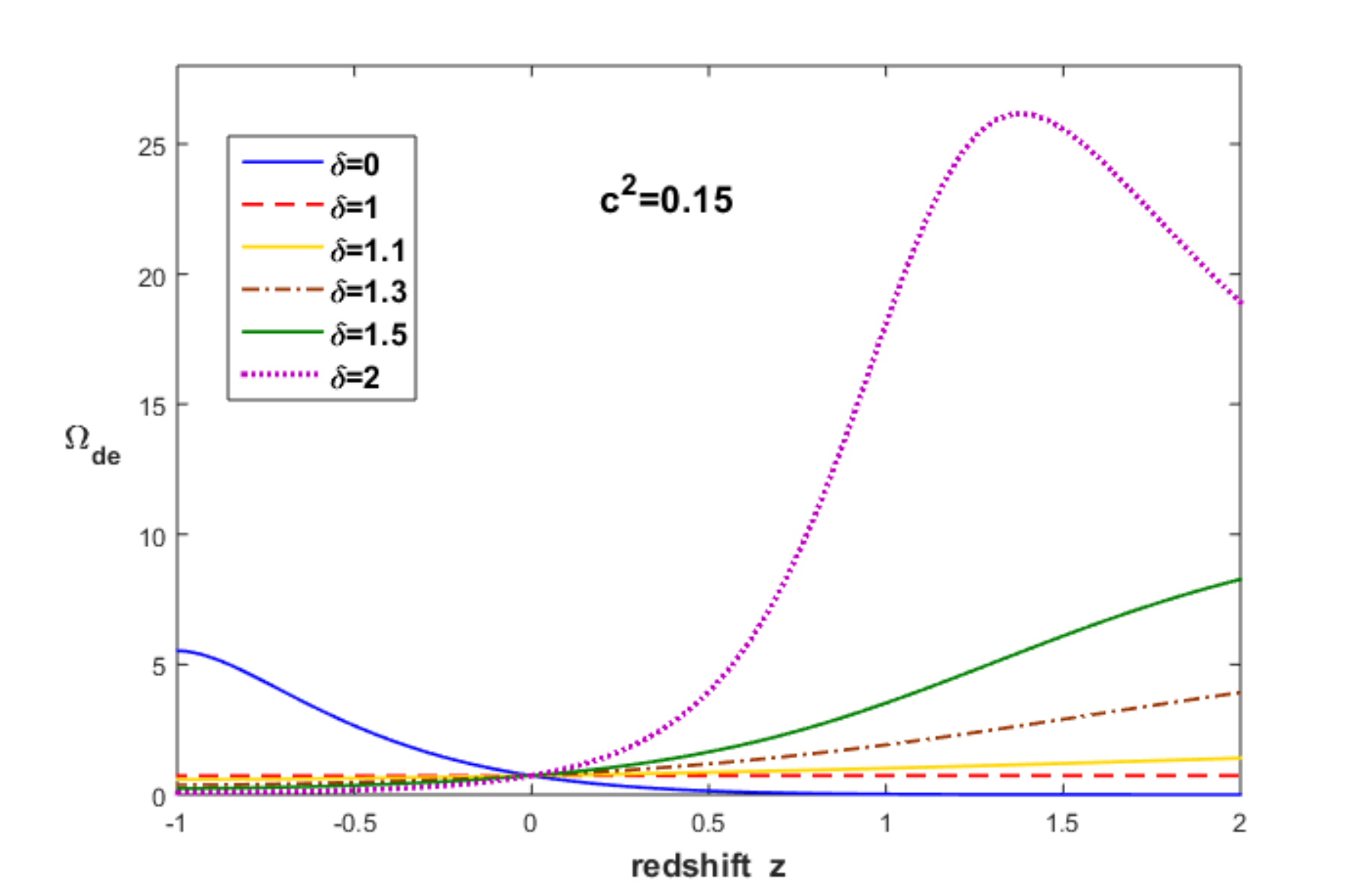}\par 
    \includegraphics[width=1.1\linewidth]{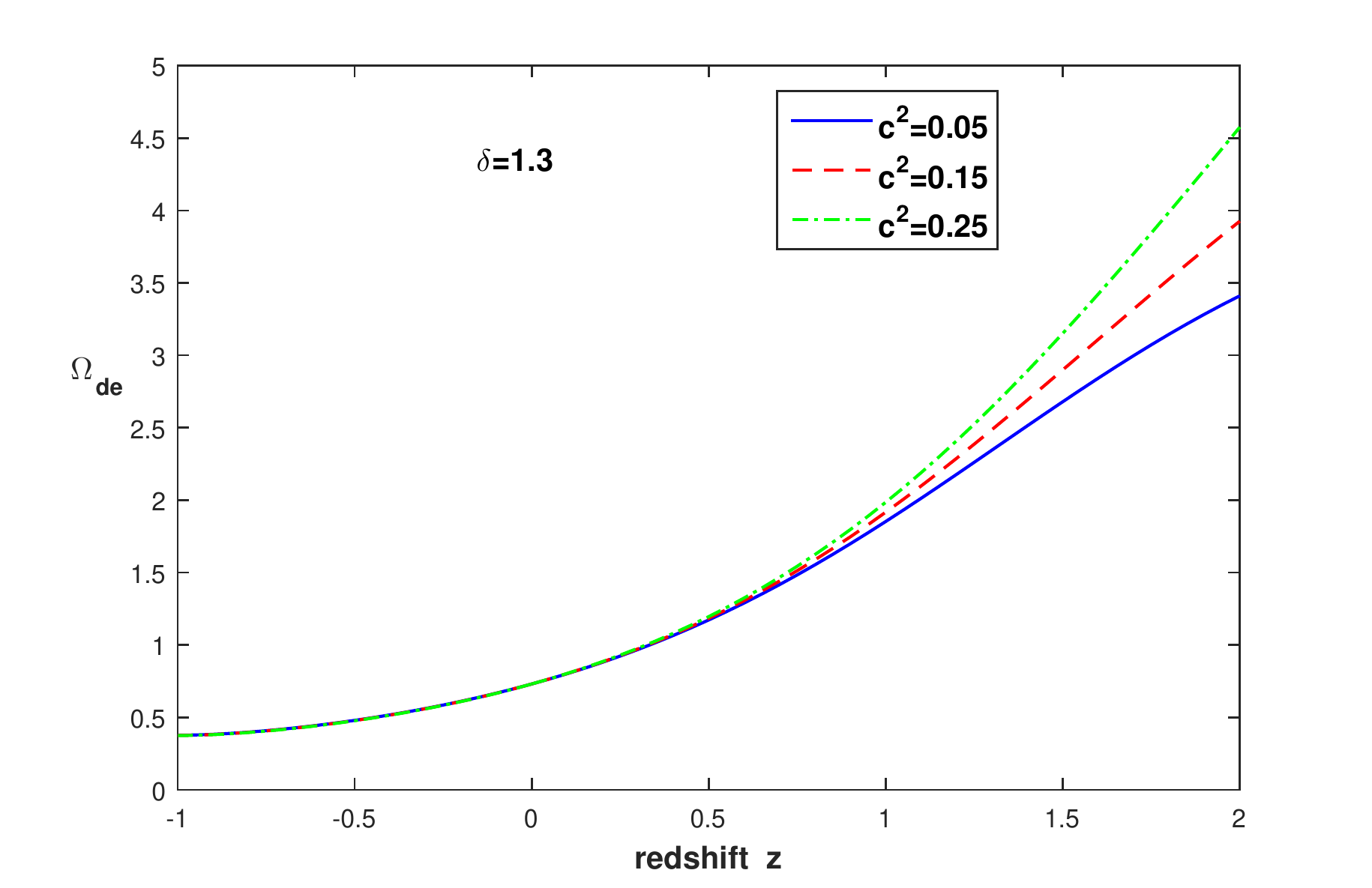}\par 
    \end{multicols}
\caption{Plot of density parameter of interacting THDE model versus redshift $z$ for $\beta_1=7.1$, $\beta_2=2.15$, $w=1000$, $\Omega^0_{de}=0.73$, $u=0.3$ and $\delta= 1.3$.}\label{fig:iom1}
\end{figure*}

We solve the differential equation \eqref{e20a} for dimensionless density parameter $\Omega_{de}$ in interacting THDE model and plot it against redshift $z$ as shown in Fig. \ref{fig:iom1}. Here, we have analyzed all the dynamical parameters through both $\delta$ and coupling coefficient $c^2$. Remaining constant parameters are same as used in previous section. It can be seen that the density parameter $\Omega_{de}$ shows decreasing behavior for all values of $\delta$ except for $\delta=0$. Also, we observe that increase of the coupling coefficient $c^2$ and parameter $\delta$ cause the $\Omega_{de}$ increases but, the effect of coupling coefficient vanishes near past and $\Omega_{de}$ becomes more steeper with the increase of $\delta$. 

From Eqs. \eqref{e14a}, \eqref{e13b} and \eqref{e20}, we find that

\begin{widetext}
\begin{eqnarray}
\label{e21}\omega_{de}&=&-1-c^2(u+1)-\frac{\beta_2a}{3(\beta_1+\beta_2 a)~\ln(\beta_1+\beta_2 a)}-\frac{2(\delta-2)}{3}\Bigg\{\bigg\{\frac{3\beta_2 a(2-\Omega_{de})}{(\beta_1+\beta_2 a)~\ln(\beta_1+\beta_2 a)}\nonumber\\&&-\frac{w\beta_2^2a^2}{(\beta_1+\beta_2 a)~[\ln(\beta_1+\beta_2 a)]^2}-\frac{3\beta_2^2 a^2}{(\beta_1+\beta_2 a)^2~\ln(\beta_1+\beta_2 a)}+\frac{w\beta_2^3a^3}{(\beta_1+\beta_2 a)^2~[\ln(\beta_1+\beta_2 a)]^2}\nonumber\\&&+\frac{w\beta_2^3a^3}{(\beta_1+\beta_2 a)^3~[\ln(\beta_1+\beta_2 a)]^3}-9\Omega_{de}\left(u(c^2-1)+c^2\right)\bigg\}\nonumber\\&&\times\bigg\{6+6(\delta-2)\Omega_{de}+\frac{6\beta_2 a}{(\beta_1+\beta_2 a)~[\ln(\beta_1+\beta_2 a)]}-\frac{w\beta_2^2a^2}{(\beta_1+\beta_2 a)~[\ln(\beta_1+\beta_2 a)]^2}\bigg\}^{-1}\Bigg\}
\end{eqnarray}
\end{widetext}

\begin{figure*}
\begin{multicols}{2}
    \includegraphics[width=1.1\linewidth]{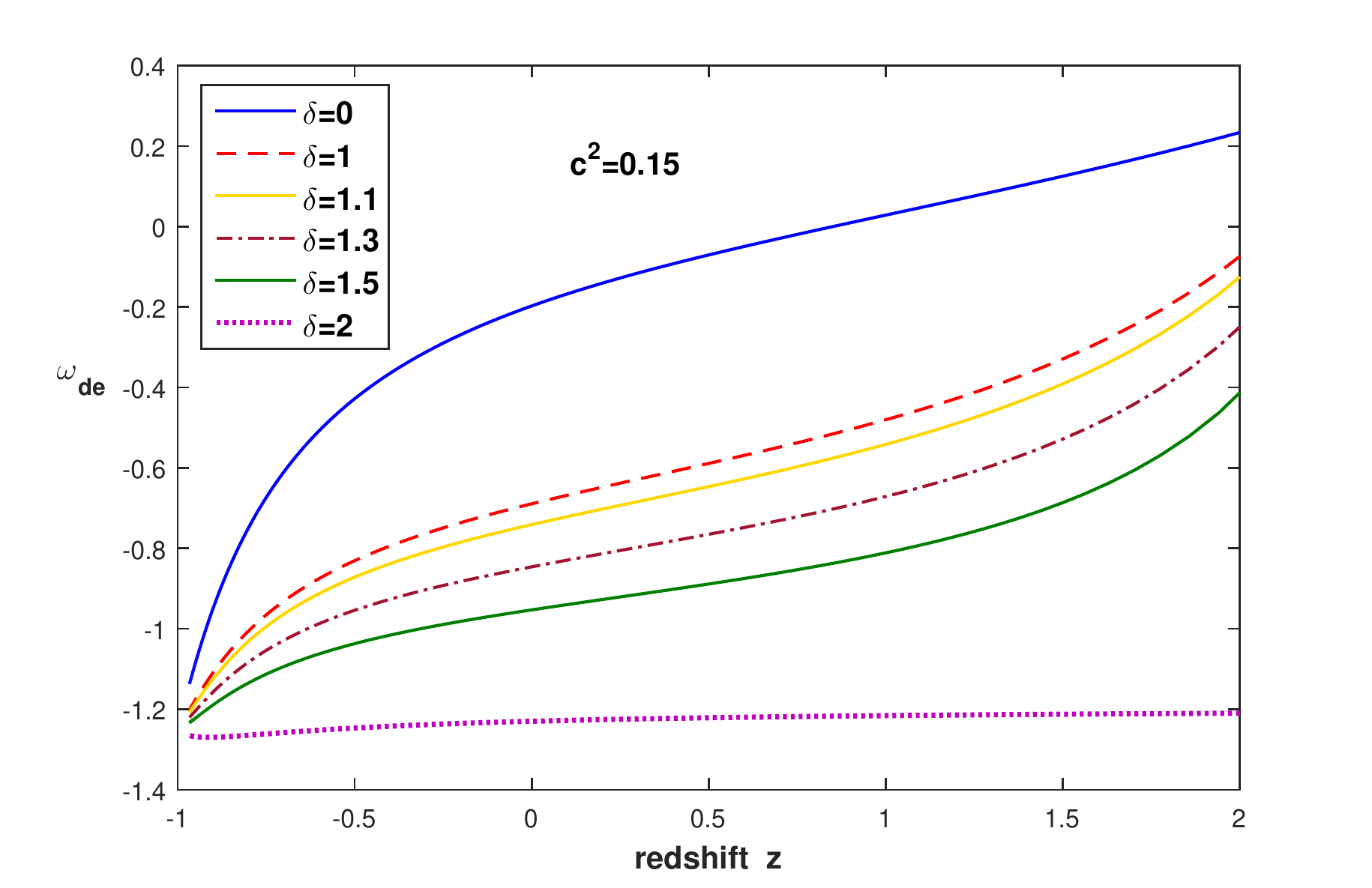}\par 
    \includegraphics[width=1.1\linewidth]{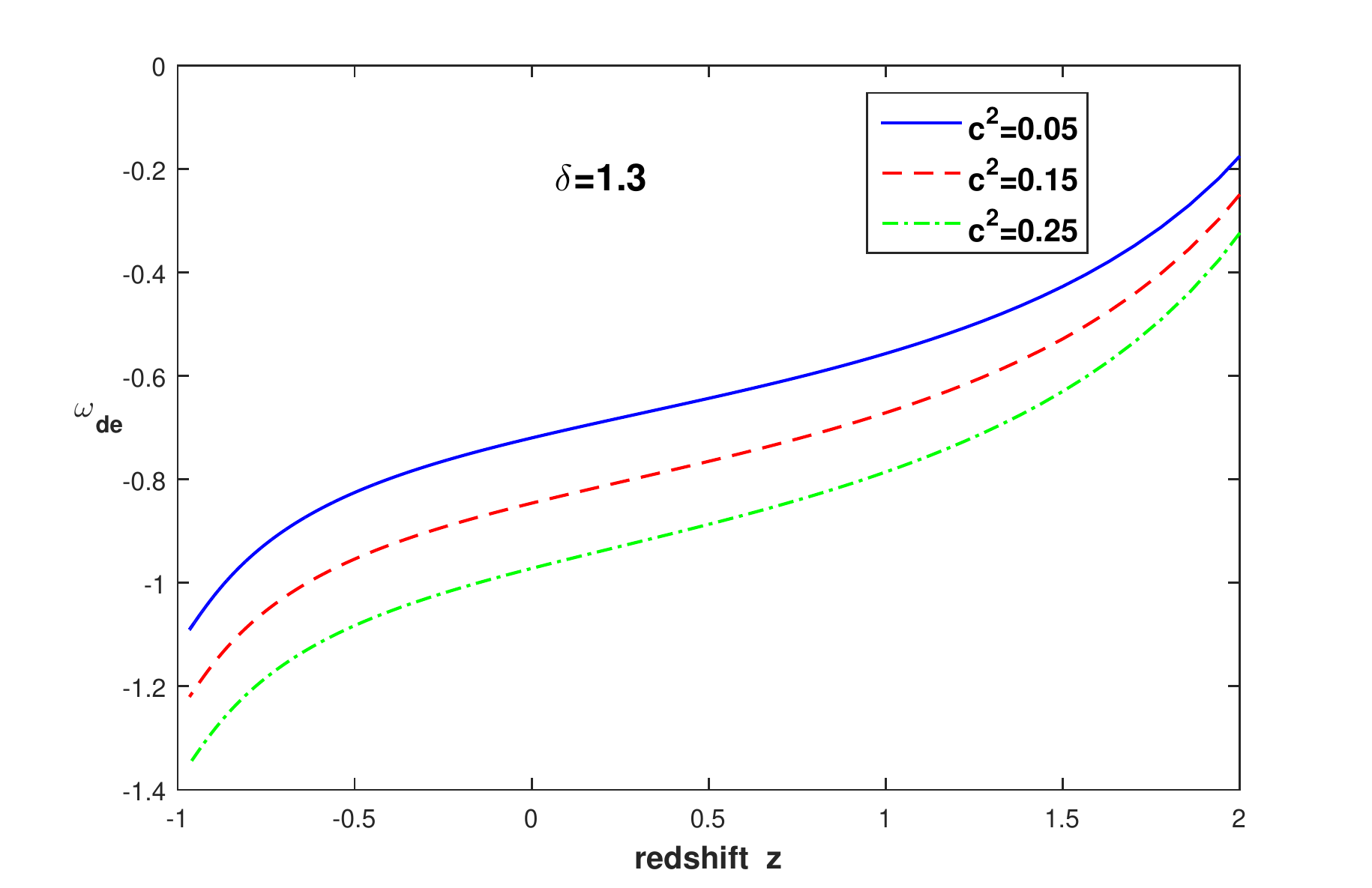}\par 
    \end{multicols}
\caption{Plot of EoS parameter of interacting THDE model versus redshift $z$ for $\beta_1=7.1$, $\beta_2=2.15$, $w=1000$, $\Omega^0_{de}=0.73$, $u=0.3$ and $\delta= 1.3$.}
	\label{fig:eos1a}
\end{figure*}

We can get EoS parameter behavior from it's plots which are shown in Fig. \ref{fig:eos1a}, for various values of $c^2$ and $\delta$. It can be observed from Fig. \ref{fig:eos1a} that the EoS parameter starts from quintessence phase and turns towards phantom region by crossing vacuum dominated era (phantom divide line $\omega_{de}=-1$) of the universe for all the values of $\delta$ and $c^2$. At late times, it can also be observed that the EoS parameter attains high phantom region with the increase of coupling coefficient $c^2$ but, attains a constant value in phantom region for various values of $\delta$. { For $\delta=0$ the model varies from matter dominated phase to dark energy phase, whereas for $\delta=2$ the model remains in the phantom region ($\omega_{de}<-1$)}. It is interesting to note here that as $\delta$ and $c^2$ increases, there is a delay in transition of the model from quintessence to phantom phase.  

The deceleration parameter is given by

\begin{widetext}
\begin{eqnarray}
\label{e22}q&=&-1+\bigg\{\frac{3\beta_2 a(2-\Omega_{de})}{(\beta_1+\beta_2 a)~\ln(\beta_1+\beta_2 a)}-\frac{w\beta_2^2a^2}{(\beta_1+\beta_2 a)~[\ln(\beta_1+\beta_2 a)]^2}-\frac{3\beta_2^2 a^2}{(\beta_1+\beta_2 a)^2~\ln(\beta_1+\beta_2 a)}\nonumber\\&&+\frac{w\beta_2^3a^3}{(\beta_1+\beta_2 a)^2~[\ln(\beta_1+\beta_2 a)]^2}+\frac{w\beta_2^3a^3}{(\beta_1+\beta_2 a)^3~[\ln(\beta_1+\beta_2a)]^3}-9\Omega_{de}\left(u(c^2-1)+c^2\right)\bigg\}\nonumber\\&&\times\bigg\{6+6(\delta-2)\Omega_{de}+\frac{6\beta_2 a}{(\beta_1+\beta_2 a)~[\ln(\beta_1+\beta_2 a)]}-\frac{w\beta_2^2a^2}{(\beta_1+\beta_2 a)~[\ln(\beta_1+\beta_2 a)]^2}\bigg\}^{-1}
\end{eqnarray}
\end{widetext}

\begin{figure*}
\begin{multicols}{2}
    \includegraphics[width=1.1\linewidth]{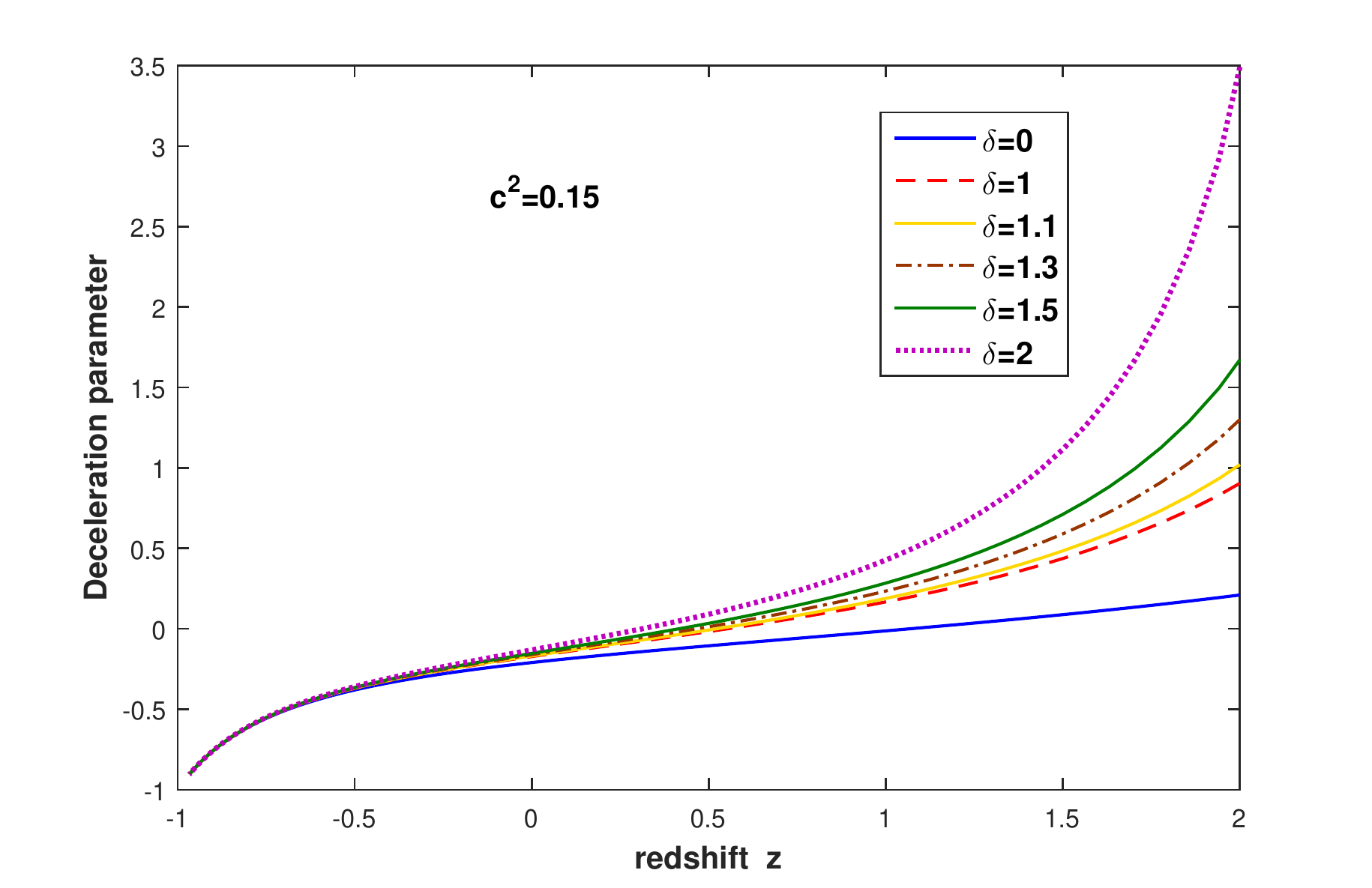}\par 
    \includegraphics[width=1.1\linewidth]{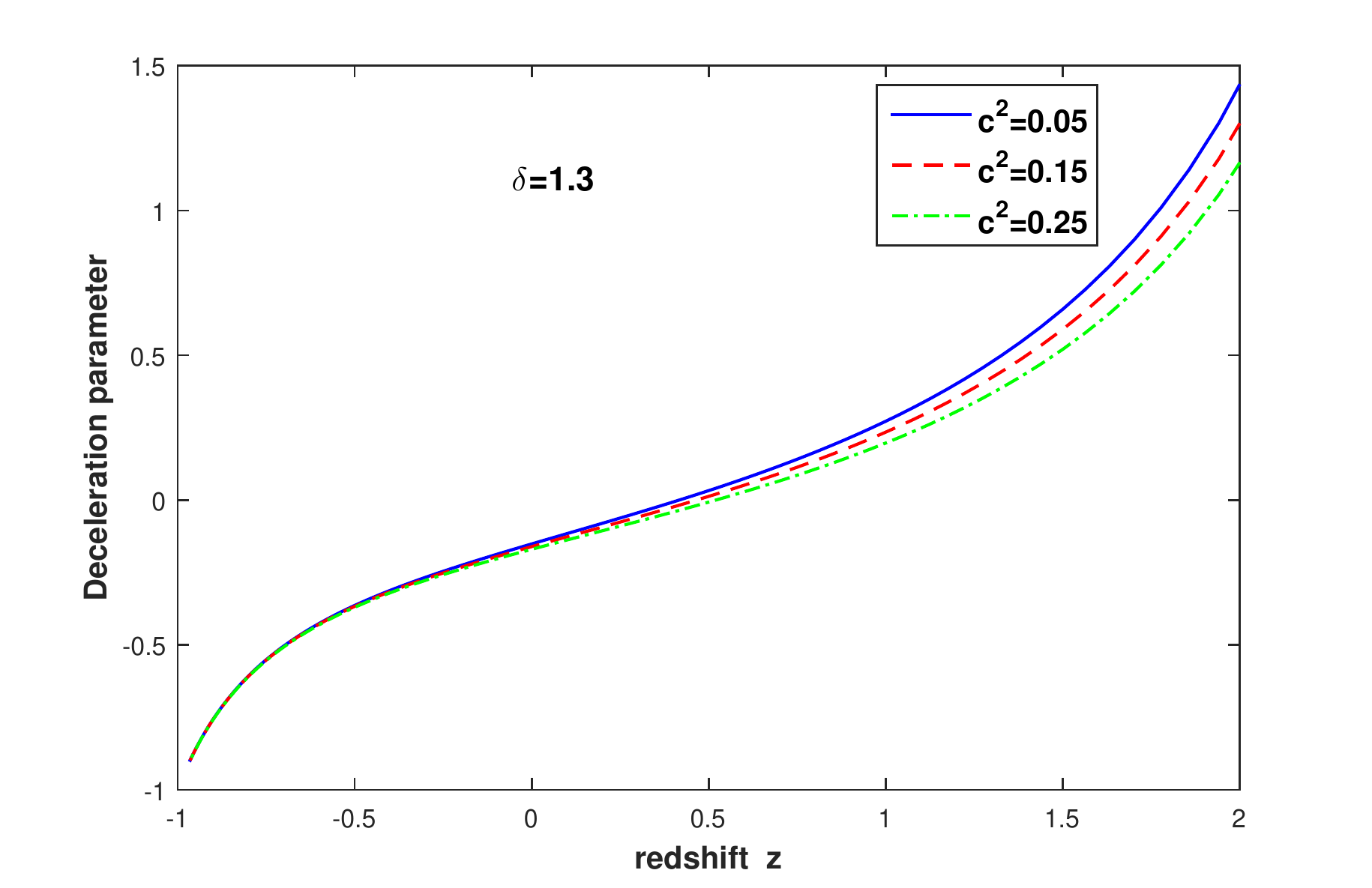}\par 
    \end{multicols}
\caption{Plot of deceleration parameter of of interacting THDE model versus redshift $z$ for $\beta_1=7.1$, $\beta_2=2.15$, $w=1000$, $\Omega^0_{de}=0.73$ and $u=0.3$.}
			\label{fig:nq1a}
\end{figure*}

Fig. \ref{fig:nq1a} represents the behavior of deceleration parameter for interacting THDE model for different values of $\delta$ and $c^2$. In both cases, it shows that the deceleration parameter starts from decelerating phase, then goes towards accelerating phase and finally approaches to $q=-1$. Also, the transition from deceleration to acceleration occurred between $z=0.6$ to $0.8$. It may be noted here that the transition redshift of the model decreases as $\delta$ increases and it is quite opposite in case of coupling coefficient $c^2$, i.e., the transition is delayed as coupling coefficient increases. 

\begin{figure*}
\begin{multicols}{2}
    \includegraphics[width=1\linewidth]{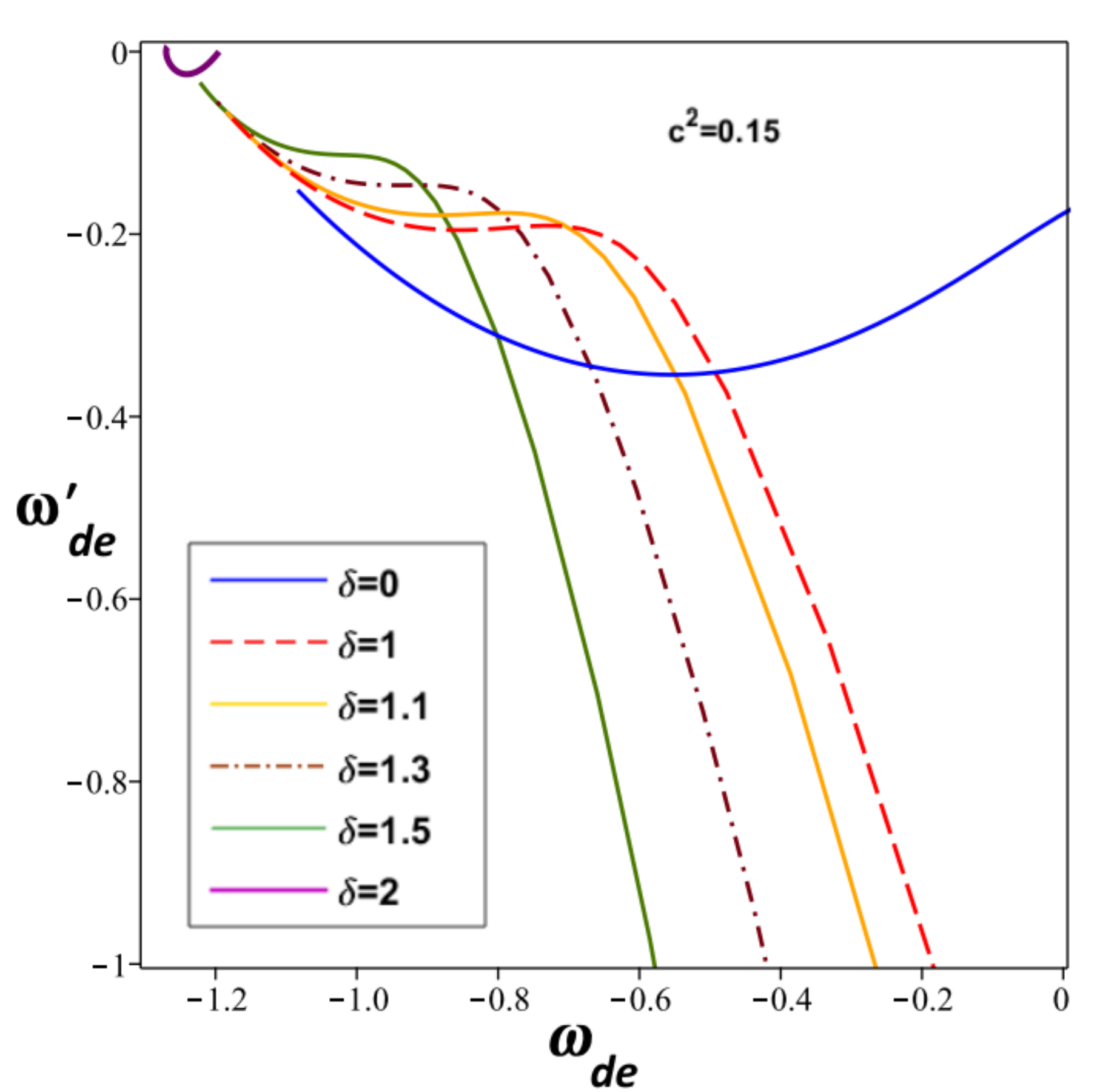}\par 
    \includegraphics[width=1\linewidth]{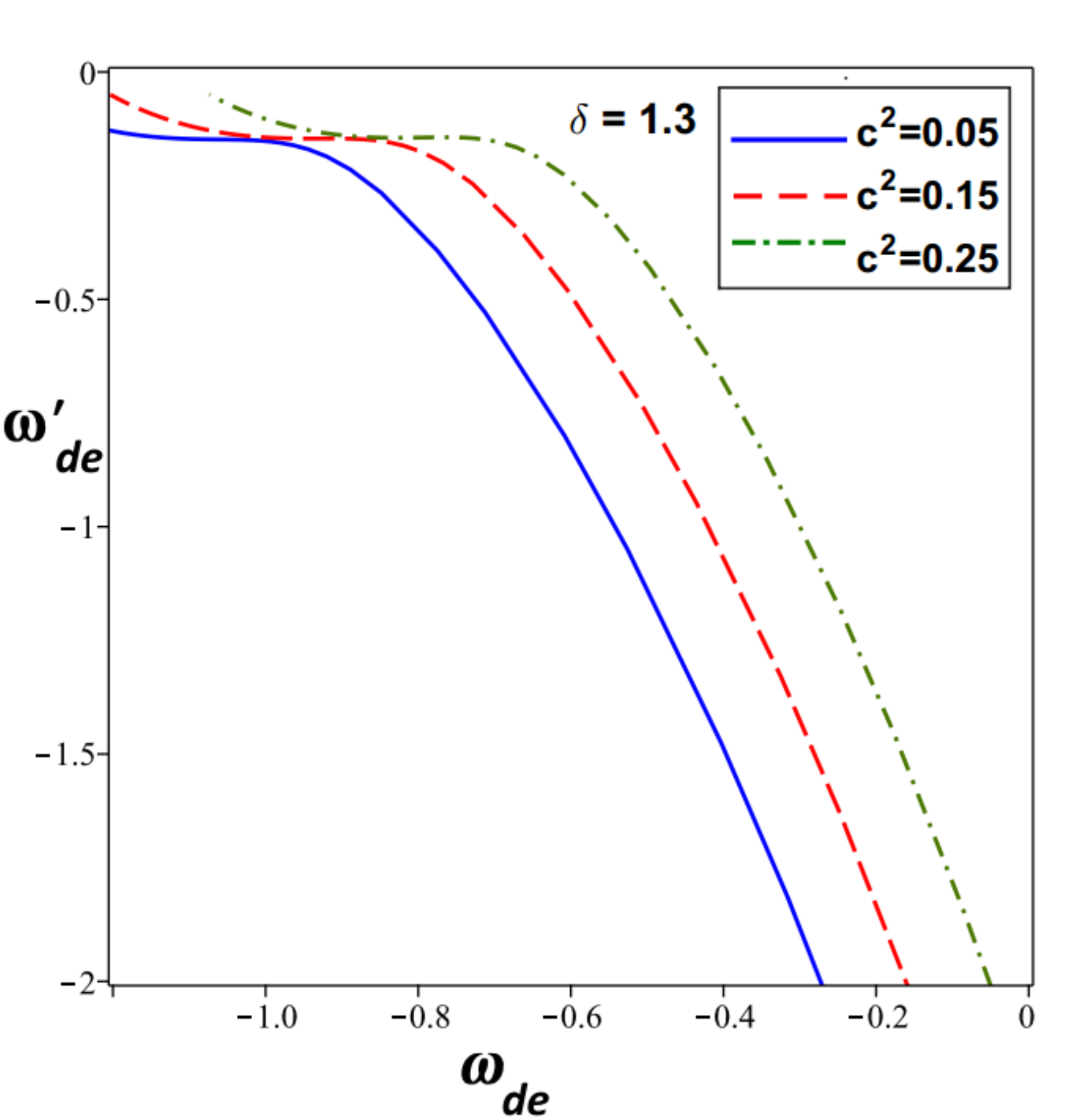}\par 
    \end{multicols}
\caption{Plot of $\omega_{de}-\omega_{de}^\prime$ plane of interacting THDE model for $\beta_1=7.1$, $\beta_2=2.15$, $w=1000$, $\Omega^0_{de}=0.73$, $u=0.3$ and $c^2=0.15$.}
\label{fig:ieospa}
\end{figure*}

The $\omega_{de}-\omega_{de}^\prime$ plane is used to present the dynamical property of dark energy models, where $\omega_{de}^\prime$ is the evolutionary form of $\omega_{de}$ with prime indicates derivative with respect to $\ln$ $a$. The $\omega_{de}-\omega_{de}^\prime$ plane for the constructed interacting THDE model is developed for various values of $\delta$ and $c^2$ as shown in Fig. \ref{fig:ieospa}. We observed from these plots that $\omega_{de}-\omega_{de}^\prime$ plane corresponds to the freezing region only. It is concluded that $\omega_{de}-\omega_{de}^\prime$ plane analysis for the present scenario gives consistent results with the accelerated expansion of the universe.

\begin{figure*}
\begin{multicols}{2}
    \includegraphics[width=1\linewidth]{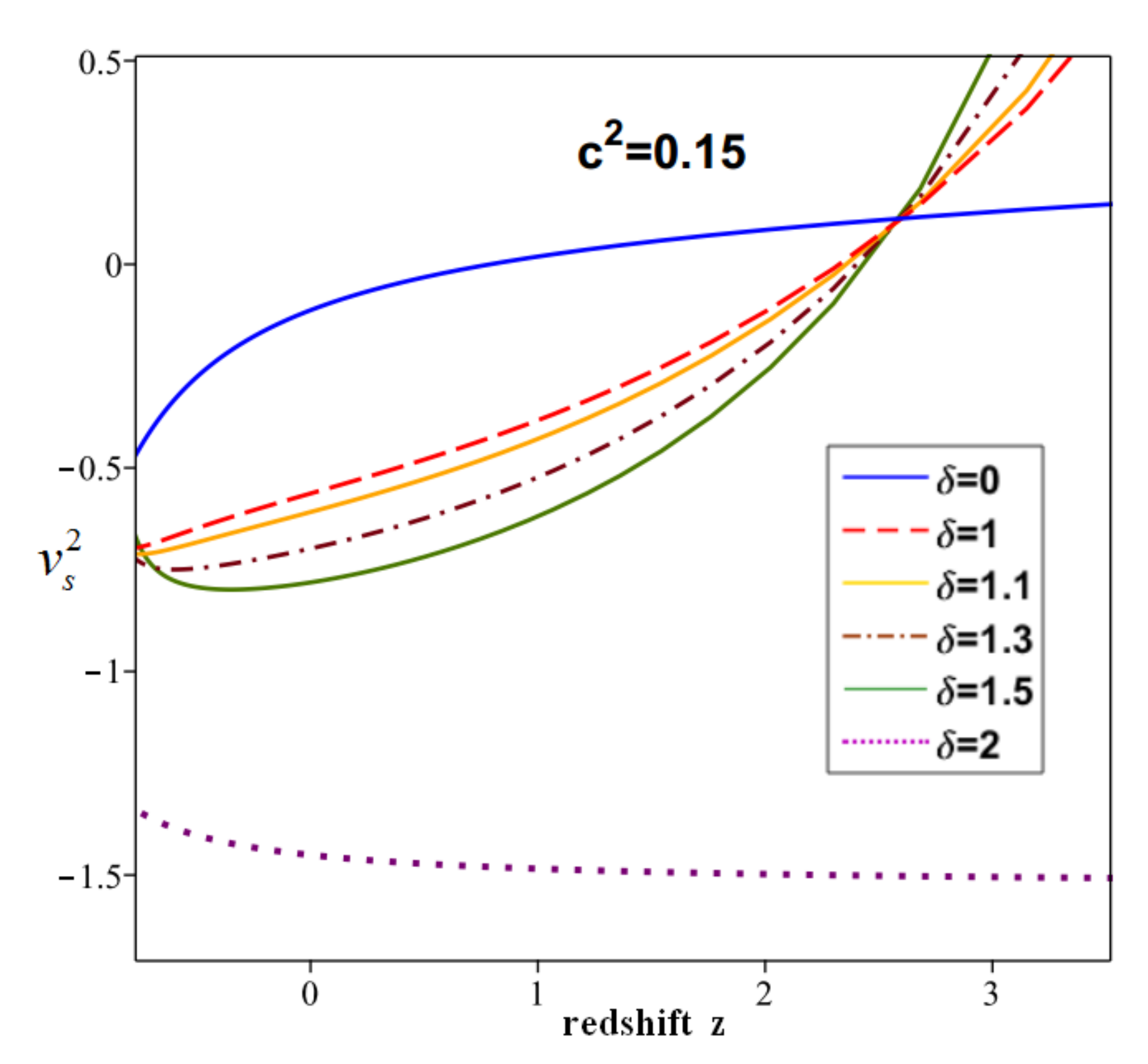}\par 
    \includegraphics[width=1\linewidth]{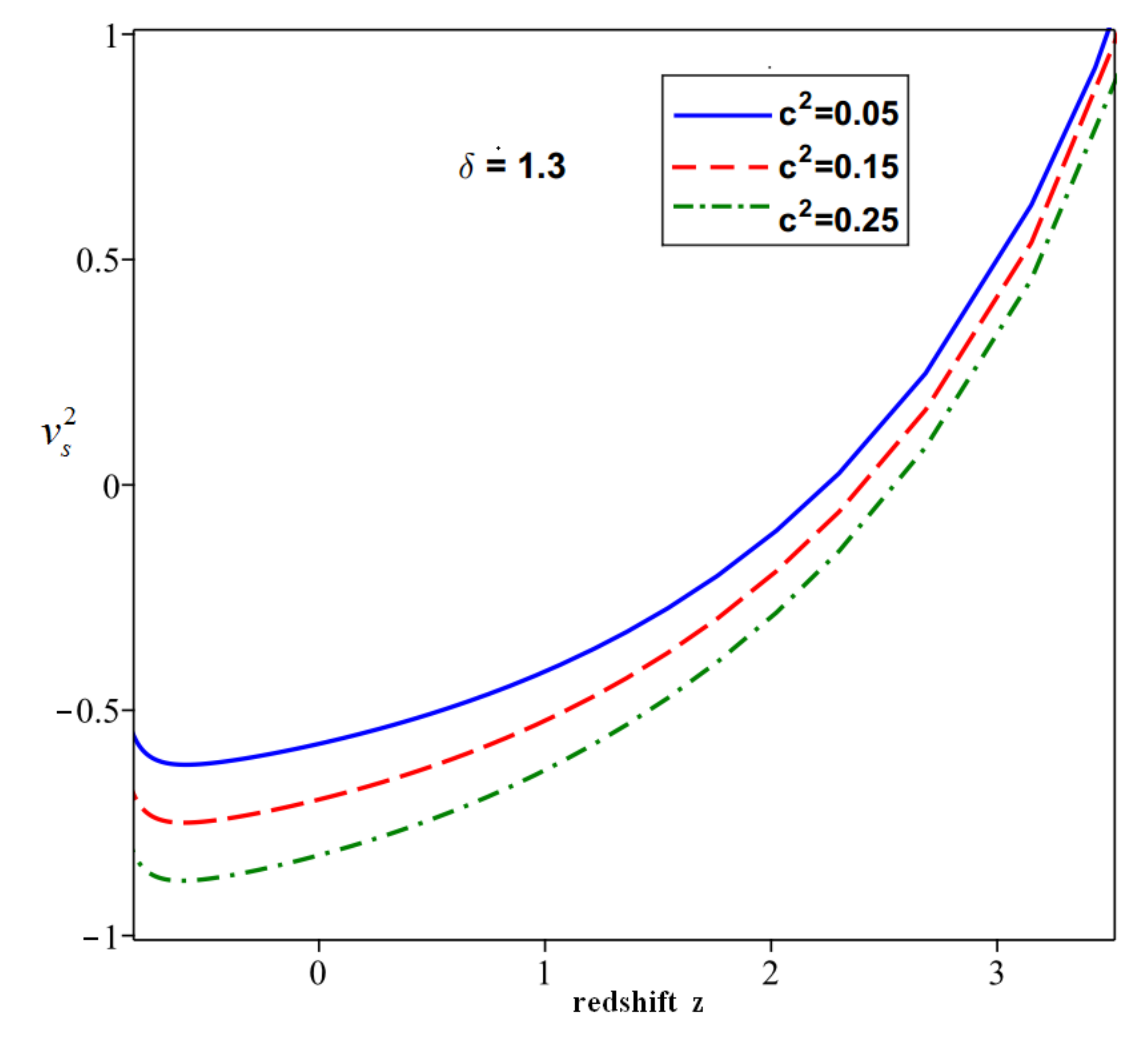}\par 
    \end{multicols}
\caption{Plot of squared sound speed of interacting THDE model versus redshift $z$ for $\beta_1=7.1$, $\beta_2=2.15$, $w=1000$, $\Omega^0_{de}=0.73$ and $u=0.3$.}
	\label{fig:ivs}
\end{figure*}

Using Eqs. \eqref{e20} and \eqref{e21} in Eq. \eqref{vs2} we can obtain the expression for squared sound speed $v_s^2$ of interacting THDE model in BD theory. In Fig. \ref{fig:ivs}, we plot its obtained expression in terms of redshift $z$ for same values of the other constants involved in the equations as for above figures. It can be seen that initially $v_s^2$ shows a decreasing behavior but with positive sign which shows the stable model. As the model evolves, these trajectories bear a negative behavior. Thus for the interacting THDE model in BD theory is stable at initial epoch and then exhibits instability at late times. 

\section{Conclusions}
\hspace*{0.6 cm} The current scenario of accelerated expansion of the universe has become more fascinating with the passage of time. In order to address this problem, different approaches have been adopted through lot of dynamical DE models and modified theories of gravity. In this work, we investigate this phenomenon by assuming the Tsallis Holographic DE within Brans-Dicke scalar-tensor theory of gravity by taking the BD scalar field as a logarithmic function of average scale factor. We have studied both scenarios i.e., interaction and non-interaction between DE and pressure less DM. We have summarized our results as follows:

$\bullet$ In order to explore the viability of constructed THDE models, we have obtained different dynamical cosmological parameters. In their evolution (non - interacting and interacting) the THDE parameter $\delta$ and coupling coefficient $c^2$ play a significant role. Hence, we have explored the dynamics of physical parameters in terms of redshift $z$ for the values of $\delta=0,~1,~1.1,~1.3,$ $~1.5,~2$ and $c^2=0.05,~0.15,~0.25$. We solve THDE density parameters (Eqs. \eqref{e12c} and \eqref{e21}), in both non-interacting and interacting models, numerically corresponding to redshift $z$ and their outputs are plotted in Figs. \ref{fig:om} and \ref{fig:iom1}. It can be seen that the $\Omega_{de}$ is positive and decreasing function throughout the evolution, and at present epoch it approaches to $0.73$ for all the three values of $\delta$ and $c^2$. Also, $\Omega_{de}$ increases with $\delta$ and $c^2$. But for $\delta=0$, the energy density parameter increases. { Planck's observations (2014,2018) \cite{ade,agh} have put the following constraints on the DE density parameter $\Omega_{de}=0.717^{+0.023}_{-0.020}$ (Planck+WP), $0.717^{+0.028}_{-0.024}$ (WMAP-9), $0.679 \pm 0.013$ (TT +lowE), $0.699 \pm 0.012$ (TE +lowE), $0.711^{+0.033}_{-0.026}$ (EE +lowE), $0.6834 \pm 0.0084$ (TT, TE, 
\\
EE+lowE), $0.6847 \pm 0.0073$ (TT, TE, EE +lowE +lensing), $0.6889\pm 0.0056$ (TT, TE, EE +lowE +lensing +BAO).} We observed that, in both non-interacting and interacting, the THDE density parameter meets the above mentioned limits which shows that our results are consistent.  

$\bullet$  The behaviors of EoS parameter versus redshift for non-interacting and interacting THDE models are respectively given in Figs. \ref{fig:eos} and \ref{fig:eos1a}. We observe that the non-interacting model starts from matter dominated era, then goes towards quintessence DE era and finally approaches to vacuum DE era for all three values of $\delta$ (Fig. \ref{fig:eos}). { For $\delta=2$, the EoS parameter becomes $-1$ i.e., cosmological constant, which is in agreement with the results obtained by Saridakis et al. \cite{sar18}. Also, we observed that as $\delta$ decreases the transition from matter dominated phase to dark energy phase is delayed is delayed considerably.} It may be noted that the non-interacting model never crosses the PDL ($\omega_{de}=-1$). Also, the EoS parameter shows translation from matter dominated era and goes towards phantom region by crossing the phantom divide line (Fig. \ref{fig:eos1a}). In interacting THDE case, the model starts from accelerated phase and then goes towards phantom DE era by crossing the quintessence era as well as PDL (i.e. vacuum DE) (Fig. \ref{fig:eos1a}). This type of behavior corresponds to quintom. It can also be observed that as $\delta$ and $c^2$ increases the transition from quintessence to phantom phase of the interacting THDE model is delayed. { The present Planck collaboration data (2018) \cite{agh} gives the limits for EoS parameter as 

{\small \begin{eqnarray}
\omega_{de}&=&-1.56^{+0.60}_{-0.48} ~ \text{(Planck + TT + lowE)}\nonumber\\
\omega_{de}&=&-1,58^{+0.52}_{-0.41} ~ \text{(Planck + TT,TE,EE+lowE)}\nonumber\\
\omega_{de}&=&-1.57^{+0.50}_{-0.40} ~ \text{(Planck + TT,TE,EE+lowE+lensing)}\nonumber\\
\omega_{de}&=&-1.04^{+0.10}_{-0.10} ~ \nonumber\\\nonumber&&\text{(Planck + TT,TE,EE+lowE+lensing+BAO)}\nonumber
\end{eqnarray}} 

It can be seen that the trajectories of EoS parameter of both interacting and non-interacting THDE models (Figs. \ref{fig:eos} and \ref{fig:eos1a}) coincide with this Planck collaboration (2018) results. }

$\bullet$  The trajectory of deceleration parameter versus redshift for non-interacting and interacting models are shown in Figs. \ref{fig:nq} and \ref{fig:nq1a}, respectively. It can be seen that, in both the cases, the model exhibits a smooth transition from early deceleration era to the current acceleration era of the universe. Also, the transition redshift $z_t$ from a deceleration to acceleration lies within the interval $0.58<z_t<0.8$. Also, the observational data from $H(z)+SN~Ia$ for $\Lambda$CDM model given a range for transition redshift as $z_t=0.682 \pm 0.082$ and $q\rightarrow -1$ as $z\rightarrow -1$. Hence, we can say that the above results are in accordance with the recent cosmological observations \cite{gio,kom}. 

$\bullet$  In this present scenario, we develop the squared speed of sound $v_s^2$ trajectories for both non-interacting and interacting THDE models (Figs. \ref{fig:vs} and \ref{fig:ivs}). In both the models, $v_s^2$ varies in positive region initially which shows the stability of the models. However, it has become negative within short interval of time and remains negative forever exhibits instability of the models later epochs. Many authors \cite{zad,tav18} in the literature have shown that the non-interacting or interacting THDE models with different IR cutoff's are unstable. However, it is interesting to mention that for increasing value of $\delta$ and decreasing value of interaction parameter $c^2$, we expect stability of the both models in near future. The $\omega_{de}-\omega_{de}^\prime$ plane for the non-interacting and interacting THDE models is also developed as given in Figs. \ref{fig:eosp} and \ref{fig:ieospa}. It may be observed from these trajectories that $\omega_{de}-\omega_{de}^\prime$ plane corresponds to freezing only. Many researchers have concluded that the expansion of the universe is comparatively more accelerating in freezing region. Also, in both the models $\omega_{de}-\omega_{de}^\prime$ plane meets the observational data from Planck collaboration (Planck + WP + BAO, \cite{ade}) which is given by $-0.89\leq\omega_{de}\leq-1.38$ and $\omega_{de}^\prime<1.32$.

Now it will be interesting to compare our THDE models in BD theory with logarithmic scalar field with the other THDE models in literature with regard to the energy density ($\Omega_{de}$), EoS ($\omega_{de}$), deceleration ($q$) parameters and the stability analysis $v_s^2$. Zadeh et al. \cite{zad} has explored the effects of different IR cutoffs on the properties of THDE model. It seems that our findings are coincide with the results of Ghaffari et al. \cite{gaf1}, Zadeh et al. \cite{zad} and Tavayef et al. \cite{tav18}. Ghaffari et al. \cite{gaf} have investigated THDE in BD theory, while Jawad et al. \cite{jaw} have studied THDE in modified BD theory with scalar field as power function of average scale factor. Ghaffari et al. \cite{gaf1} have shown that the density parameter is increasing function and converges to $\Omega_{de}=1$ at late times, but in our models (both non-interacting and interacting) the density parameter has opposite behavior, i.e., $\Omega_{de}$ is decreasing positive function and finally tends to a positive constant value. This is due to because of logarithmic BD scalar field in our model. However, the present value of $\Omega_{de}$ is same as in \cite{gaf} and also coincide with the observational results. Our THDE non-interacting model never crosses the PDL but, in the work of Jawad et al. \cite{jaw} the EoS parameter behaves like quintom. We, also, observed from the findings of Ghaffari et al. \cite{gaf} that the transition redshift of each dynamical parameter is same, but in our models it varies through $\delta$ and $c^2$. Our both THDE models are unstable for any value of $\delta$ or coupling constant $c^2$, a result the same as that of the models in  \cite{gaf,jaw}. The above results leads to the conclusion that our THDE models in BD theory with logarithmic form of scalar field are in good agreement with the observational data and also we hope that the above investigations will help to have a deep insight into the behavior of THDE universe in BD cosmology.   

It is interesting to mention here that, for $\delta=0$, the dynamical behavior of properties such as $\Omega_{de}$ (Figs. \ref{fig:om} and \ref{fig:iom1}), $\omega_{de}-\omega_{de}^\prime$ plane (Figs. \ref{fig:eosp} and \ref{fig:ieospa}) and squared sound speed (Fig. \ref{fig:ivs}) have some peculiar behavior when compared with the other values of $\delta$. For $\delta=0$, it is clear from the Figs. \ref{fig:om} and \ref{fig:iom1} that  the density parameter $\Omega_{de}$ positive and increasing function which is quite opposite to the behavior of $\Omega_{de}$ when $\delta$ is non-zero. Also, for $\delta=0$, the interacting THDE model is unstable near present epoch whereas for other non-zero values of $\delta$ the model becomes unstable near past. The interacting THDE model varies in DE region only for non-zero values of $\delta$ whereas for $\delta=0$ the interacting model varies in both matter dominated and DE regions.          
\\\\
\textbf{Acknowledgments:} We are very much grateful to the honorable referee and the editor for the illuminating suggestions that have significantly improved our work in terms of research quality and presentation. S.M. acknowledges Department of Science \& Technology (DST), Govt. of India, New Delhi,
for awarding Junior Research Fellowship \ \  (File No.
DST/INSPIRE Fellowship/2018/IF180676). PKS acknowledges CSIR, New Delhi, India for financial support to carry out the Research project [No.03(1454)/19/EMR-II Dt.02/08/2019].

\end{document}